\documentclass[preprint]{aastex631}

\usepackage{natbib}
\usepackage{amsmath}

\newcommand{\kms}{\mbox{km\,s$^{-1}$}}

\newcommand{\Msun}{\mbox{M$_{\odot}$}}

\newcommand{\E}[1]{\hbox{$10^{ #1 }$}}

\def\HII{\hbox{H \small{II}}}

\begin{document}

\title{MagMar III - Resisting the Pressure, Is the Magnetic Field Overwhelmed in NGC6334I?}
\shortauthors{Cort\'es et al.}

\shortauthors{Cort\'es et al.}

\author[0000-0002-3583-780X]{Paulo C. Cort\'es}
\affiliation{Joint ALMA Observatory, Alonso de C\'ordova 3107, Vitacura, Santiago, Chile}
\affiliation{National Radio Astronomy Observatory, 520 Edgemont Road, Charlottesville, VA 22903, USA}

\author[0000-0002-3829-5591]{Josep M. Girart}
\affiliation{Institut de Ci\'encies de l'Espai (ICE-CSIC), Campus UAB, Carrer de Can Magrans S/N, E-08193 Cerdanyola del Vall\'es, Catalonia}
\affiliation{Institut d'Estudis Espacials de Catalunya (IEEC), E-08034 Barcelona, Catalonia}

\author[0000-0002-7125-7685]{Patricio Sanhueza}
\affiliation{National Astronomical Observatory of Japan, 2-21-1 Osawa, Mitaka, Tokyo 181-8588, Japan}
\affiliation{Department of Astronomical Science, SOKENDAI (The Graduate University for Advanced Studies), 2-21-1 Osawa, Mitaka, Tokyo 181-8588, Japan}
\author[0000-0002-4774-2998]{Junhao Liu}
\affiliation{National Astronomical Observatory of Japan, 2-21-1 Osawa, Mitaka, Tokyo 181-8588, Japan}

\author[0000-0001-9281-2919]{Sergio Mart\'in}
\affiliation{European Southern Observatory, Alonso de C\'ordova 3107, Vitacura, Santiago, Chile}
\affiliation{Joint ALMA Observatory, Alonso de C\'ordova 3107, Vitacura, Santiago, Chile}
\author[0000-0003-3017-4418]{Ian W. Stephens}
\affiliation{Department of Earth, Environment and Physics, Worcester State University, Worcester, MA 01602, USA}
\author[0000-0002-1700-090X]{Henrik Beuther}
\affiliation{Max Planck Institute for Astronomy, K\"onigstuhl 17, 69117 Heidelberg, Germany}

\author[0000-0003-2777-5861]{Patrick M. Koch}
\affiliation{Institute of Astronomy and Astrophysics, Academia Sinica, 11F of Astronomy-Mathematics Building, AS/NTU No.1, Sec. 4, Roosevelt Rd, Taipei 10617, Taiwan,
Republic of China}

\author[0000-0001-5811-0454]{M. Fern\'andez-L\'opez}
\affiliation{Instituto Argentino de Radioastronom\'\i a (CCT-La Plata, CONICET; CICPBA; UNLP), C.C. No. 5, 1894, Villa Elisa, Buenos Aires, Argentina}

\author[0000-0002-3078-9482]{\'Alvaro S\'anchez-Monge}
\affiliation{Institut de Ci\`encies de l'Espai (ICE, CSIC), Carrer de Can Magrans s/n, E-08193, Bellaterra, Barcelona, Spain}
\affiliation{Institut d'Estudis Espacials de Catalunya (IEEC), E-08034, Barcelona, Spain}
\author[0000-0002-6668-974X]{Jia-Wei Wang}
\affiliation{Academia Sinica Institute of Astronomy and Astrophysics, No.1, Sec. 4., Roosevelt Road, Taipei 10617, Taiwan}
\author[0000-0002-6752-6061]{Kaho Morii}
\affiliation{Department of Astronomy, Graduate School of Science, The University of Tokyo, 7-3-1 Hongo, Bunkyo-ku, Tokyo 113-0033, Japan}
\affiliation{National Astronomical Observatory of Japan, National Institutes of Natural Sciences, 2-21-1 Osawa, Mitaka, Tokyo 181-8588, Japan}
\author[0000-0003-1275-5251]{Shanghuo Li }
\affiliation{Max Planck Institute for Astronomy, K\"onigstuhl 17, 69117 Heidelberg, Germany}
\author[0000-0002-0028-1354]{Piyali Saha}
\affiliation{National Astronomical Observatory of Japan, National Institutes of Natural Sciences, 2-21-1 Osawa, Mitaka, Tokyo 181-8588, Japan}
\author{Qizhou Zhang}
\affiliation{Harvard-Smithsonian Center for Astrophysics, 60 Garden Street, Cambridge, MA 02138, USA}

\author{David Rebolledo}
\affiliation{Joint ALMA Observatory, Alonso de C\'ordova 3107, Vitacura, Santiago, Chile}
\affiliation{National Radio Astronomy Observatory, 520 Edgemont Road, Charlottesville, VA 22903, USA}

\author{Luis A. Zapata}
\affiliation{Instituto de Radioastronom\'ia  y Astrof\'isica, Universidad Nacional Autónoma de M\'exico, P.O. Box 3-72, 58090, Morelia, Michoac\'an, M\'exico}
\author[0000-0001-7379-6263]{Ji-hyun Kang}
\affiliation{Korea Astronomy and Space Science Institute (KASI), 776 Daedeokdae-ro, Yuseong-gu, Daejeon 34055, Republic of Korea}

\author[0000-0001-9822-7817]{Wenyu Jiao}
\affiliation{Kavli Institute for Astronomy and Astrophysics, Peking
University, Haidian District, Beijing 100871, People’s Republic of
China}
\affiliation{Department of Astronomy, School of Physics, Peking
University, Beijing, 100871, People's Republic of China}

\author[0000-0002-1229-0426]{Jongsoo Kim}
\affiliation{Korea Astronomy and Space Science Institute (KASI), 776 Daedeokdae-ro, Yuseong-gu, Daejeon 34055, Republic of Korea}

\author[0000-0002-0786-7307]{Yu Cheng}
\affiliation{National Astronomical Observatory of Japan, 2-21-1 Osawa, Mitaka, Tokyo 181-8588, Japan}

\author[0000-0001-7866-2686]{Jihye Hwang}
\affiliation{Korea Astronomy and Space Science Institute (KASI), 776 Daedeokdae-ro, Yuseong-gu, Daejeon 34055, Republic of Korea}

\author[0000-0003-0014-1527]{Eun Jung Chung}
\affiliation{Department of Astronomy and Space Science, Chungnam National University, Daejeon, Republic of Korea}

\author[0000-0002-7497-2713]{Spandan Choudhury}
\affiliation{Korea Astronomy and Space Science Institute (KASI), 776 Daedeokdae-ro, Yuseong-gu, Daejeon 34055, Republic of Korea}

\author[0000-0002-9907-8427]{A-Ran Lyo}
\affiliation{Korea Astronomy and Space Science Institute (KASI), 776 Daedeokdae-ro, Yuseong-gu, Daejeon 34055, Republic of Korea}

\author[0000-0002-8250-6827]{Fernando Olguin}
\affiliation{Institute of Astronomy and Department of Physics, National Tsing Hua University, Hsinchu 30013, Taiwan}

\email{paulo.cortes@alma.cl}

\begin{abstract}
We report on ALMA observations of polarized dust emission at 1.2 mm from NGC6334I, a source known for its significant flux outbursts. Between five months, our data show no substantial change in total intensity  and a modest 8\% variation in linear polarization, suggesting a phase of stability or the conclusion of the outburst. The magnetic field, inferred from this polarized emission, displays a predominantly radial pattern from North-West to South-East with intricate disturbances across major cores, hinting at spiral structures. Energy analysis of CS$(J=5 \rightarrow 4)$ emission yields an outflow energy of approximately $3.5\times10^{45}$ ergs, aligning with previous interferometric studies. Utilizing the Davis-Chandrasekhar-Fermi method, we determined magnetic field strengths ranging from 1 to 11 mG, averaging at 1.9 mG. This average increases to 4 $\pm 1$ mG when incorporating Zeeman measurements. Comparative analyses using gravitational, thermal, and kinetic energy maps reveal that magnetic energy is significantly weaker, possibly explaining the observed field morphology. 
	We also find that the energy in the outflows and the expanding cometary {\HII} region is also larger than the magnetic energy, suggesing that protostellar feedback maybe the dominant driver behind the injection of turbulence in NGC6334I at the scales sampled
by our data.
The gas in NGC6334I predominantly exhibits supersonic and trans-Alfvenic conditions, transitioning towards a super-Alfvenic regime, underscoring a diminished influence of the magnetic field with increasing gas density. These observations are in agreement with prior polarization studies at 220 GHz, enriching our understanding of the dynamic processes in high-mass star-forming regions.
\end{abstract}


\section{INTRODUCTION}\label{se:intro}

Stars are formed within dense and weakly ionized gas and dust complexes commonly known as molecular clouds, where the current paradigm suggests that they are filamentary in nature \citep{Arzoumanian2011}.
Despite their weakly ionization rates, 
magnetic fields thread these regions and are unavoidable. 
In the past decades, the advent of millimeter and submillimeter interferometers such as Berkeley, Illinois and Maryland Association (BIMA), the Combined Array for Research in Millimeter-wave Astronomy (CARMA), and Sub-Millimeter Array (SMA) began to allow surveys of magnetic fields in a sample of star forming regions \citep{Rao1998, Cortes2005, Hull2014, Zhang2014}.
The arrival of the Atacama Large millimeter/sub-millimeter Array (ALMA) and the James Clerk-Maxwell Telescope JMCT-Pol2, further improve the understanding of the effects of magnetic fields in star forming regions \citep{Pattle2023}. However, a complete understanding of their effects  is still elusive.
This is particularly evident when studying the formation of high-mass stars (over 8 \Msun) because of their relative faraway distances and the complexity inherent to the high-mass star formation process.
Through the detection of polarized emission from dust, where the main assumption is that dust-grains are 
align by magnetic fields, detailed projected magnetic field morphology maps have been uncovered by ALMA from number of high-mass star forming regions \citep[e.g. ][]{Cortes2016, Beltran2019,Liu2020,Sanhueza2021,Fernandez-Lopez2021,Cortes2021a,Cortes2021b,Beuther2023,Liu2023a,Liu2023b}. 
Here, we delve into the magnetic environment surrounding one of these high-mass star-forming clumps, NGC6334I, as part of the Magnetic fields in Massive star-forming Regions (MagMar) collaboration \citep{Sanhueza2024}. NGC6334I is a part of the vast Giant Molecular Cloud (GMC) NGC6334, situated within the Sagittarius-Carina spiral arm, approximately $1.3 \pm 0.3$ kpc away from the Sun \citep{Chibueze2014}. Despite NGC6334 vastness ($\sim 100$ pc), the regions of NGC6334I and NGC6334I(N) concentrate most of the studies in the millimeter and sub-millimeter range of the spectrum \citep{McCutcheon2000,Hunter2014,Hunter2017,Sadaghiani2020}. A notable characteristic of NGC6334I is a significant millimeter flux outburst, predominantly centered on the MM1B proto-stellar core \citep{Brogan2016A, Hunter2017, Hunter2018, Brogan2018}. This flux surge was accompanied by a flare in multiple maser species, particularly the OH maser transition which offered insights into line-of-sight (LoS) magnetic field strengths \citep{MacLeod2018}. Moreover, several methanol maser lines, some previously unobserved, flared around the MM1 proto-cluster, hinting at a possible accretion event.
Furthermore, outflow emission is also observed throughout NGC6334I. Initially detected in CO by \citet{Bachiller1990} using the IRAM 30 meter telescope, outflow emission was later observed in CS by \citet{McCutcheon2000} using the JCMT, and by \citet{Beuther2008} in HCN using the Australian Compact Array (ATCA). Later,  ALMA observations of CS$(J=6 \rightarrow 5)$  spatially resolved the emission finding four bipolar outflows where the most prominent one appears to originate from the MM1 proto-cluster \citep{Brogan2018}.
Finally, the magnetic field in NGC6334I has been mapped by the SMA \citep{Zhang2014a,Zhang2014b,Li2015}, the JCMT POL2 instrument at $14^{\prime \prime}$ resolution \citep{Arzoumanian2021} and by ALMA at $\sim$ 1.4 mm (220 GHz) and  0$^{\prime \prime}.6$ resolution \citep{Liu2023b}.
Here, we present 1.2 mm (250 GHz) and 0$^{\prime \prime}.4$ resolution mapping of the magnetic field in NGC6334I.
 The ensuing sections are structured as follows: Section 2 outlines the observation techniques, calibration, and data analysis methods; Section 3 unveils the results from the polarized dust  emission; Section 4 presents the discussion. We encapsulate our findings and conclusions in Section 5.

\section{OBSERVATIONS}\label{se:obs}

NGC6334I was observed as part of project 2018.1.00105.S, which was executed twice in session mode  \citep[see chapter 8 in ][ for details about the session observing mode]{Cortes2023}, on December 13$^{\mathrm{th}}$ 2018 and on May 4$^{\mathrm{th}}$ 2019 under configuration C43-4 (providing baseline lengths from 15 to 783 m).
The correlator was configured to yield full polarization cross correlations ($XX, XY, YX,$  and $YY$) using the Frequency Division Mode  (FDM), including spectral windows to map the dust continuum and windows centered on a number of  molecular line rotational transitions relevant to the study of high-mass star formation (see Table \ref{table:setup}), with an effective wavelength of 1.2 mm. 
The bandpass was calibrated using J1427-4206 for session 1 and J1924-2914 for session 2. The time dependant gain and the instrumental polarization terms were calibrated using J1717-3342 and J1751+0939, respectively.
For calibration we used CASA version 5.4 and version 5.6 for imaging \citep{CASA2022}.
To image the continuum, we manually extracted the line-free channels from each spectral window,
which we later phase-only self-calibrated using a final solution interval of 90 seconds. 
These solutions were then applied to all of the molecular line transitions presented here before imaging. The spectral cubes were produced by  using 2 {\kms} channel width.  The statistics of the flat Stokes images, before debiasing, for both continuum and channel maps are shown in Table \ref{table:stkStats}.
All of the Stokes parameters were imaged independently using the CASA task {\em tclean}, which yielded an angular resolution of approximately $0.5^{\prime \prime} \times 0.3^{\prime \prime}$, with  a position angle of -78$^{\circ}$.
The data were primary beam corrected and debiased pixel-by-pixel following \citet{Wardle1974} and \citet{Hull2015}.
Finally, we also obtained archival data from project 2017.1.00793.S which also observed NGC6334I, but using a lower frequency
spectral setup with an effective wavelength of 1.3 mm \citep[see ][ and Table \ref{table:setup}]{Liu2023}.  These data were also obtained in two sessions on June 28$^{\mathrm{th}}$ 2018 and September 2$^{\mathrm{nd}}$ 2018, roughly two months apart.
We did not combined these datasets, but briefly compared the polarized dust emission (see section \ref{se:res} and appendix \ref{ap:fcomp}).

\setlength{\tabcolsep}{2pt}
\begin{deluxetable*}{cccccccccccccccccc}
\tablecolumns{18}
\tablewidth{0pt}
\tabletypesize{\scriptsize}
\tablecaption{ Spectral Setup\label{table:setup}}
\tablehead{
\colhead{Band} &
\colhead{Spw 1} &
\colhead{$\Delta \nu_{1}$} &
\colhead{Spw 2} &
\colhead{$\Delta \nu_{2}$} &
\colhead{Spw 3} &
\colhead{$\Delta \nu_{3}$} &
\colhead{Spw 4} &
\colhead{$\Delta \nu_{4}$} &
\colhead{Spw 5} &
\colhead{$\Delta \nu_{5}$} &
\colhead{Spw 6} &
\colhead{$\Delta \nu_{6}$} &
\colhead{Spw 7} &
\colhead{$\Delta \nu_{7}$} &
\colhead{$B_\textrm{maj}$} &
\colhead{$B_\textrm{min}$} &
\colhead{PA} \\
\colhead{}     &
\colhead{(GHz)}     &
\colhead{(kHz)}     &
\colhead{(GHz)}     &
\colhead{(kHz)}     &
\colhead{(GHz)}     &
\colhead{(kHz)}     &
\colhead{(GHz)}     &
\colhead{(kHz)}     &
\colhead{(GHz)}     &
\colhead{(kHz)}     &
\colhead{(GHz)}     &
\colhead{(kHz)}     &
\colhead{(GHz)}     &
\colhead{(kHz)}     &
\colhead{($^{\prime \prime}$)}  &
\colhead{($^{\prime \prime}$)}  &
\colhead{(\arcdeg)}
}
\startdata
6 & 261.245 & 244 & 260.277 & 244.141 & 257.523  & 976 & 245.520 & 976 & 243.520 & 976 & - & - & - & - & 0.45 & 0.4 & -82.8 \\
6 & 216.430 & 976 & 218.54 & 976 & 233.54 & 976 & 230.528 & 122 & 231.051 & 122 & 231.211 & 122 & 231.312 & 122 & 1.7 & 1.1 & 87 \\
6 & 216.430 & 976 & 218.54 & 976 & 233.54 & 976 & 230.528 & 122 & 231.051 & 122 & 231.211 & 122 & 231.312 & 122 & 0.7 & 0.5 & 68
\enddata
\vspace{0.1cm}
\tablecomments{The spectral configuration and beam sizes from the MagMaR project (row 1) are detailed here along with the configuration used by project 2017.1.00793.S (rows 2 amd 3). For each spectral window (e.g. Spw 1), the center frequency is indicated in GHz, along with the channel width in kHz (e.g. $\Delta \nu_{1}$).  The  synthesized beam parameters (angular resolution) for both projects are the major axis $B_\textrm{maj}$, minor axis $B_\textrm{min}$, and position angle PA. Note, 
project 2017.1.00793.S was observed in two different configurations hence the two different beam sizes between row 2 and 3.}
\end{deluxetable*}

\setlength{\tabcolsep}{2pt}
\begin{deluxetable*}{cccc}
\tablecolumns{4}
\tablewidth{0pt}
\tabletypesize{\scriptsize}
\tablecaption{ Continuum Images Statistics\label{table:stkStats}}
\tablehead{
\colhead{Stokes} &
\colhead{Max} &
\colhead{Mean} &
\colhead{$\sigma$} \\
\colhead{}     &
\colhead{($\mathrm{mJy\,beam}^{-1}$)} &
\colhead{($\mathrm{mJy\,beam}^{-1}$)} &
\colhead{($\mu\mathrm{Jy\,beam}^{-1}$)}
}
\startdata
I & 1300 & 33.0 & 998 \\
Q & 4.1 & -0.01 & 55 \\
U & 4.1 & \phantom{0}0.07 & 55 \\
V & 1.6 & \phantom{0}0.03 & 76 
\enddata
\vspace{0.1cm}
\tablecomments{The Stokes Images statistics, from the combined sessions, are listed here. Note, Stokes $I$ is dynamic range limited; also, Stokes $Q$, $U$, and $V$ can be negative and thus, the maximum is not necessary a positive value. Furthermore, note that the units of Stokes $Q$, $U$, and $V$ are $\sigma$ are $\mu\mathrm{Jy\,beam}^{-1}$. The  Stokes $V$ values from ALMA observations are subject to larger uncertainties than linear polarization \citep[see ][]{Cortes2023}. }
\end{deluxetable*}

\section{RESULTS}
\label{se:res}

\subsection{The Dust Continuum Emission}
\begin{figure*} 
\includegraphics[width=0.95\hsize]{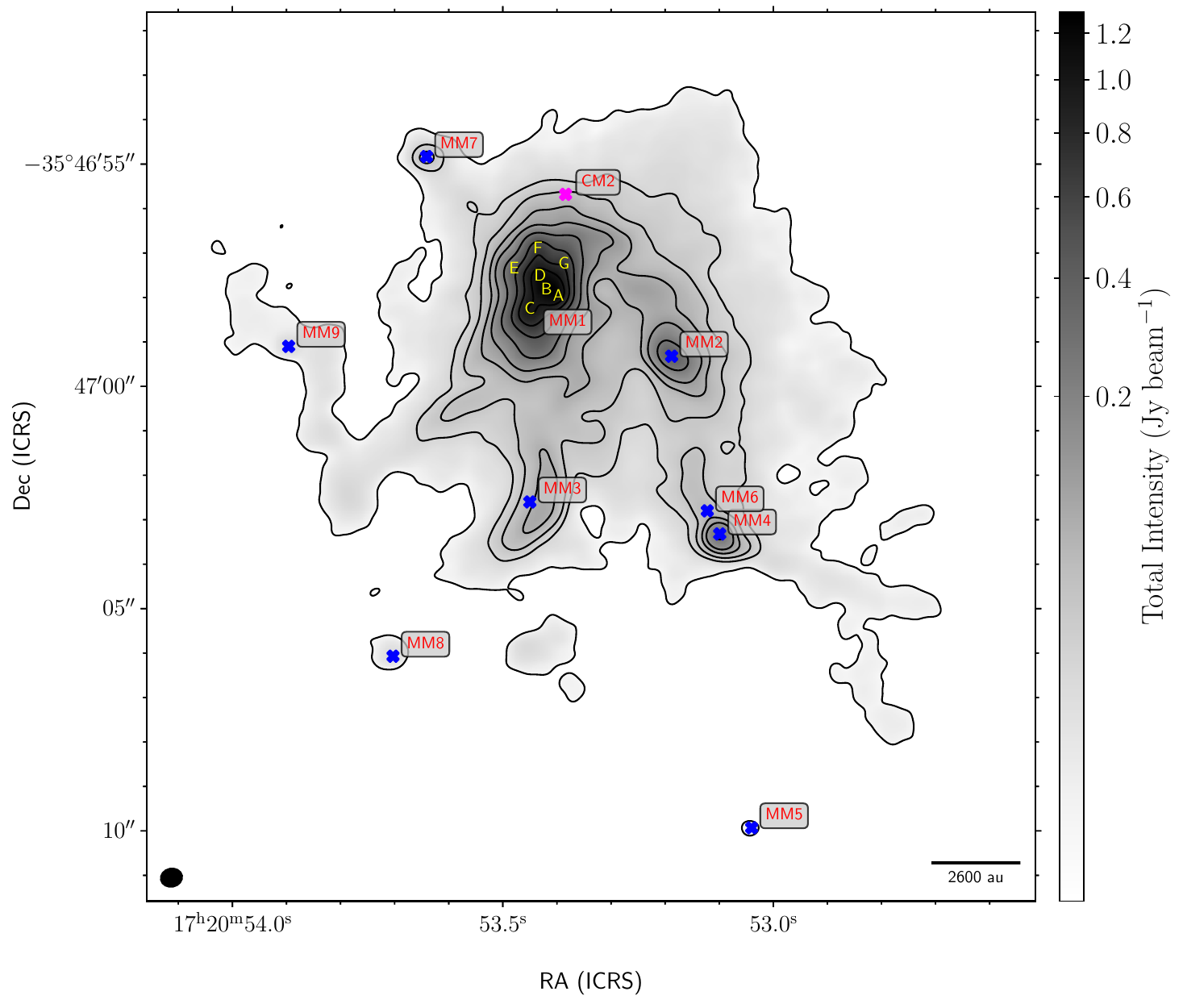}
\smallskip
\caption{This figure displays the 1.2 mm total intensity dust emission from NGC6334I. The grayscale indicates the intensity in Jy beam$^{-1}$, corresponding to the scale bar on the right. Contours represent intensity levels at 3.6, 18, 36, 72, 120, 240, 480, 720, and 960 ,mJy,beam$^{-1}$. Sources within our field of view, previously detected by \citet{Brogan2016A}, are marked with blue crosses and labeled in red. Notably, the MM1 source has shown signs of fragmentation, potentially forming a proto-cluster. The letters A through G denote the MM1 peak regions as identified by high-resolution ALMA observations. The magenta cross shows the position of the CM2 synchrotron/maser source, the bar at the bottom right corresponds a length scale of 2600 au, and the beam size of our ALMA image using robust=0.5 is shown by the black ellipse to the bottom left.
\label{fig:NGC6334I_I}
}
\bigskip
\end{figure*}

Figure \ref{fig:NGC6334I_I} shows the NGC6334I total intensity (Stokes I) map from the ALMA 1.2 mm continuum data. The most massive sources 
previously identified by \citet{Brogan2016A} are detected and are indicated by blue crosses in the map. 
The  dust continuum map at 250 GHz is also consistent with the lower frequency data obtained by \citet{Liu2023b} 
(see appendix \ref{ap:fcomp}).
The massive proto-cluster MM1 is not resolved in our data; nonetheless, we have superposed the individual peaks identified by \citet{Brogan2016A} following their nomenclature and position. The unusual source CM2, a strong water maser source with an ambiguous origin and also detected in radio continuum, is also seen in our data at the $30\sigma$ level in dust emission, where $\sigma = 1.0$ mJy beam$^{-1}$ \citep[see Table \ref{table:stkStats} and Tables in ][]{Brogan2016A}. 
The UC {\HII} region from the MM3 core is well constrained which give us confidence that most of the emission at 250 GHz is coming from dust grains.

Since our data were acquired by executing two sessions separated by roughly five months, we explored the flux variability  reported by \citet{Hunter2021}, and references therein. To this effect, we imaged each session independently in all Stokes. We found that the differences between the flux in the MM1 proto-cluster was below 1\%, below the ALMA accuracy of band 6 ($\sim$ 10\%). Because the 2017.1.00793 data (220 GHz) were taken on different configurations and only $\sim$ 2 months apart, comparing the two is difficult therefore we omit it here on. We also explored variability in the polarized flux where we found variability in Stokes $Q$ and $U$ in the order of 8\% (see Table \ref{table:fluxVar}). Note, the minimum amount of detectable linear polarization by ALMA is estimated to be 0.1\%, as established from point source measurements \citep{Cortes2023}. Because the flux variability in Stokes $I$ during this period was negligible, this suggests that the outburst might have already finished or the emission is below our sensitivity. Although a variability of 8\% in $Q$ and $U$ within six months appears intriguing, its analysis is out of the scope of this work and we defer it to a future investigation.

\subsection{Column Density and Mass}
\label{se:dust}

\begin{figure*}
\includegraphics[width=0.51\hsize]{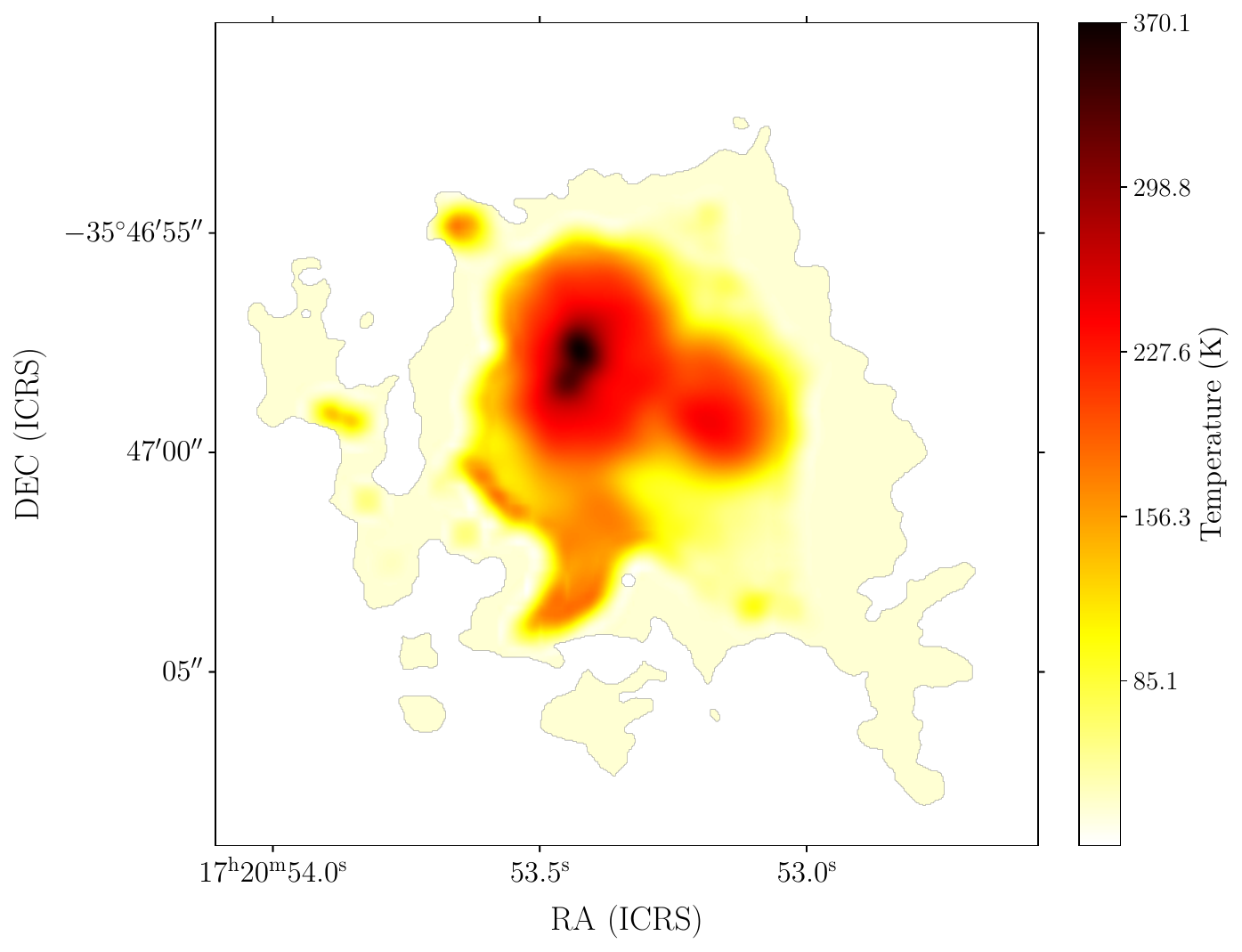}
\includegraphics[width=0.5\hsize]{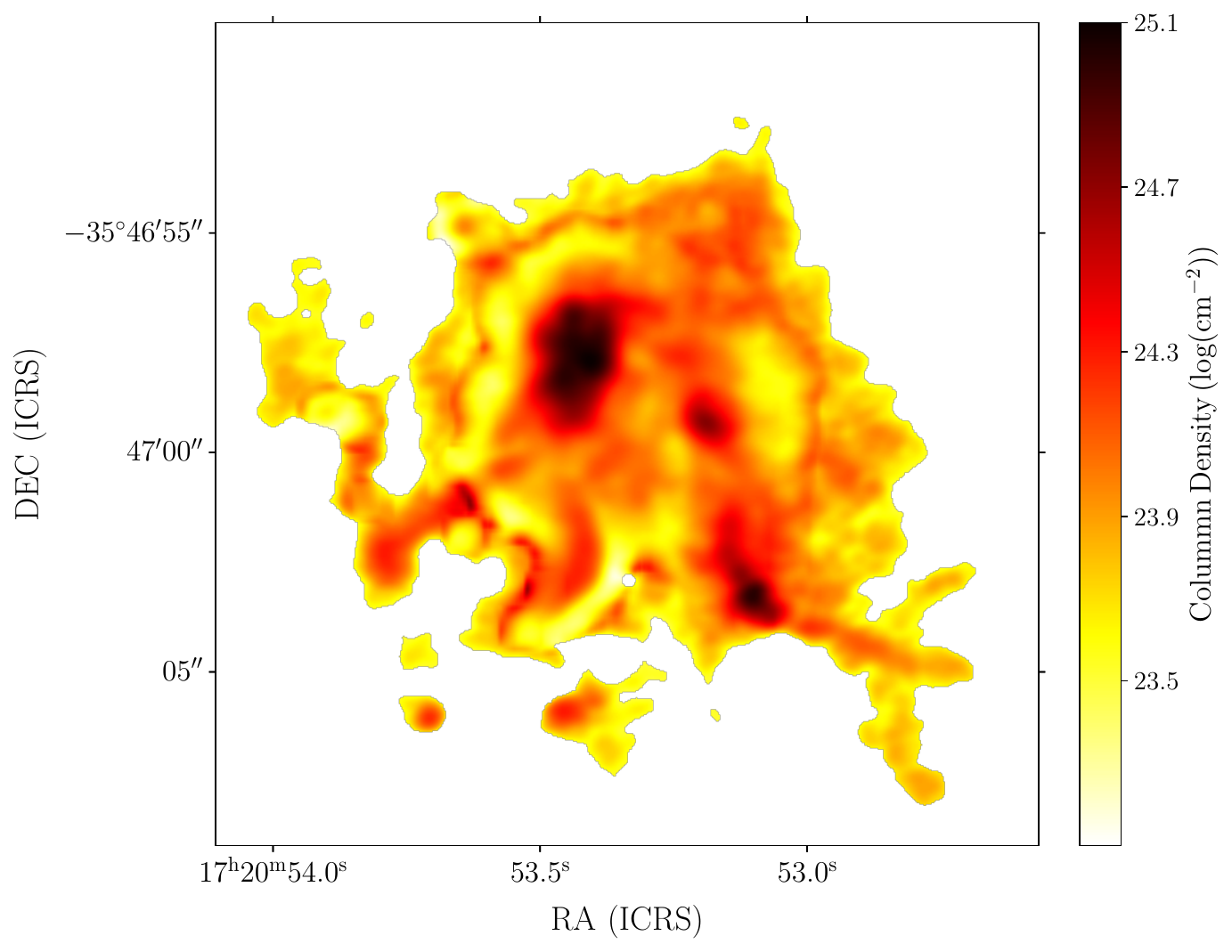}\\
\includegraphics[width=0.99\hsize]{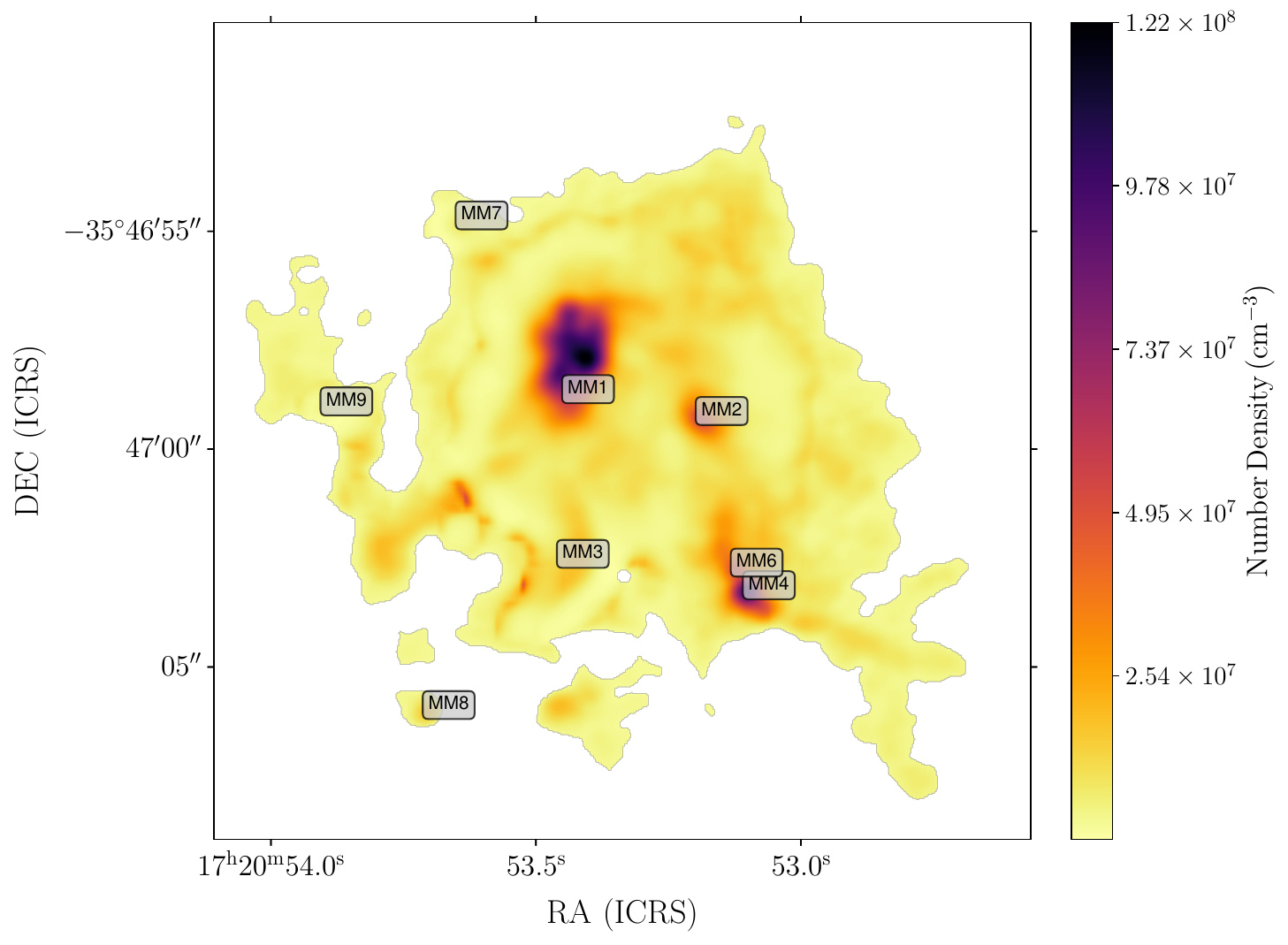}
\smallskip
\caption{{\bf \em Top)} The extended temperature model map from \citep{Liu2023} is shown here in color scale indicating dust the temperature throughout NGC6334I along with column density map in logarithmic scale derived by using the temperature model map. {\bf \em Bottom)} The number density map derived by assuming cylindrical geometry and a thickness of 0.03 pc. The main sources identified by \citet{Brogan2016A} are indicated in the map with their respective labels.
\label{fig:colDen}
}
\bigskip
\end{figure*}

Calculating the column density from dust emission at millimeter wavelength is a well established procedure.  Here, we use the standard formulation developed by \citet{Hildebrand1983} but we make use of the NGC6334I temperature model obtained by \citet{Liu2023b} from Methanol emission where we assume that the gas emission is optically thin and that the gas and dust 
are in thermal equilibrium  (see Figure \ref{fig:colDen}). The temperature model map was
extended to match the dust emission spatial distribution detected by ALMA to the $3\sigma$ level. To extend the model temperature map, we assumed that the dust temperature converged to 30 K
outside the original model map. This is  justified given the results from \citet{McCutcheon2000,Sandell2000} who derived a dust temperature value of 30 K for NGC6334I using all available millimeter and millimeter data at the time. The map was later smoothed  using a Savitzky-Golay smoothing filter \citep{SavitzkyGolay1964} with a window of one beam to ensure continuity\footnote{The Savitzky-Golay filter is a digital filter that can be applied to a set of digital data points for the purpose of smoothing the data. This is achieved by fitting successive subsets of adjacent data points with a low-degree polynomial by the method of linear least squares. We use this filter to smooth the temperature map because the filter is particularly useful to preserve important features of the data, such as relative maxima, minima, and width, which are often "washed out" by other types of smoothing filters.}. 
Besides the temperature model, we used a dust emissivity $\kappa_{\nu} = 1$ cm$^{2}$ gr$^{-1}$
for the dust grains at 230 GHz \citep{Ossenkopf1994}, and a gas to dust ratio of a 100 and the assumption that the dust emission is optically thin. 
The column density map in log scale is shown at Figure \ref{fig:colDen}. 
Note, because the temperature model derived by \citet{Liu2023b} may have used optically thick CH$_{3}$OH emission, we may be overestimating the temperature in the most dense regions of the clump, which would lead to under-estimations of the column density.
Nonetheless, the column density values at the core positions in our map appear to be in agreement with \citet{Brogan2016A} who observed NGC6334I with ALMA and the VLA computing Spectral Energy Distributions (SED) for the major sources in the region deriving temperatures and densities. 
By having a column density map\footnote{Note, we are deriving here a magnetic field map which has values per pixel. However, the limiting resolution factor is given by the synthesized beam of the telescope which should be taken into account when deriving conclusions from these maps. }, we can estimate a map of the mass. We do this by computing $\mathrm{m}_{i,j}=\mu \mathrm{m_{H}}\mathrm{N}_{i,j}\mathrm{A}$, where $i,j$ are map pixel indexes along right ascension and declination respectively, $\mathrm{A}$ is the area of each pixel, $\mathrm{m_{H}}$ is the mass of Hydrogen, $\mu = 2.8$ is the mean molecular weight, and $\mathrm{N}_{i,j}$ is the column density per pixel. 

\subsubsection{The Number Density Map}

Because of the complicated morphological features seen in the dust emission, it is difficult to model the spatial distribution of the dust and determine its volume density. A simple approach is to assume that the whole region is a cylinder of a certain thickness which can be estimated from the dust emission intensity profile. 
Opting for a cylindrical rather than spherical geometry could be more appropriate for several reasons. While a spherical model encompassing all dust emission at the 3$\sigma$ level indicates a mean radius of approximately 0.02 pc, leading to average and peak number densities significantly lower than those observed in ALMA studies of the region, a cylindrical model offers more flexibility. By adjusting the cylinder's height, it becomes possible to align the modeled core densities more closely with the observed densities ranging between $10^{7} - 10^{10}$ cm$^{-3}$, as reported by \citet{Brogan2016A,Sadaghiani2020}. This approach allows for a more accurate representation of the physical conditions within these dense cores.
Note, this simple geometrical assumption assumes the same height throughout NGC6334I which is not accurate. It is more likely that the cores in this region will follow a power law density profile different from the rest of the surrounding gas. Thus, we choose the height of the cylinder looking for a drop in the intensity as a function of distance from the peak emission and assume that this length corresponds to the height. The number density per pixel is determined then by computing $\mathrm{n}_{i,j} = \mathrm{N}_{i,j} / Z$, where $Z$ is the thickness of the cylinder. By doing this, we find a drop in the intensity at the $5\sigma$ level in the dust emission which corresponds to $\sim 5^{\prime\prime}$ or 0.03 pc at a distance of 1300 pc. 
The values presented in our number density map align with those established by \citet{Brogan2016A}, which were derived from Spectral Energy Density (SED) fitting to ascertain dust temperatures and under the assumption of spherical geometry.

\subsection{The Magnetic Field Morphology from Polarized Dust Emission}
\label{sse:fmorph}

\begin{figure*} 
\includegraphics[width=0.95\hsize]{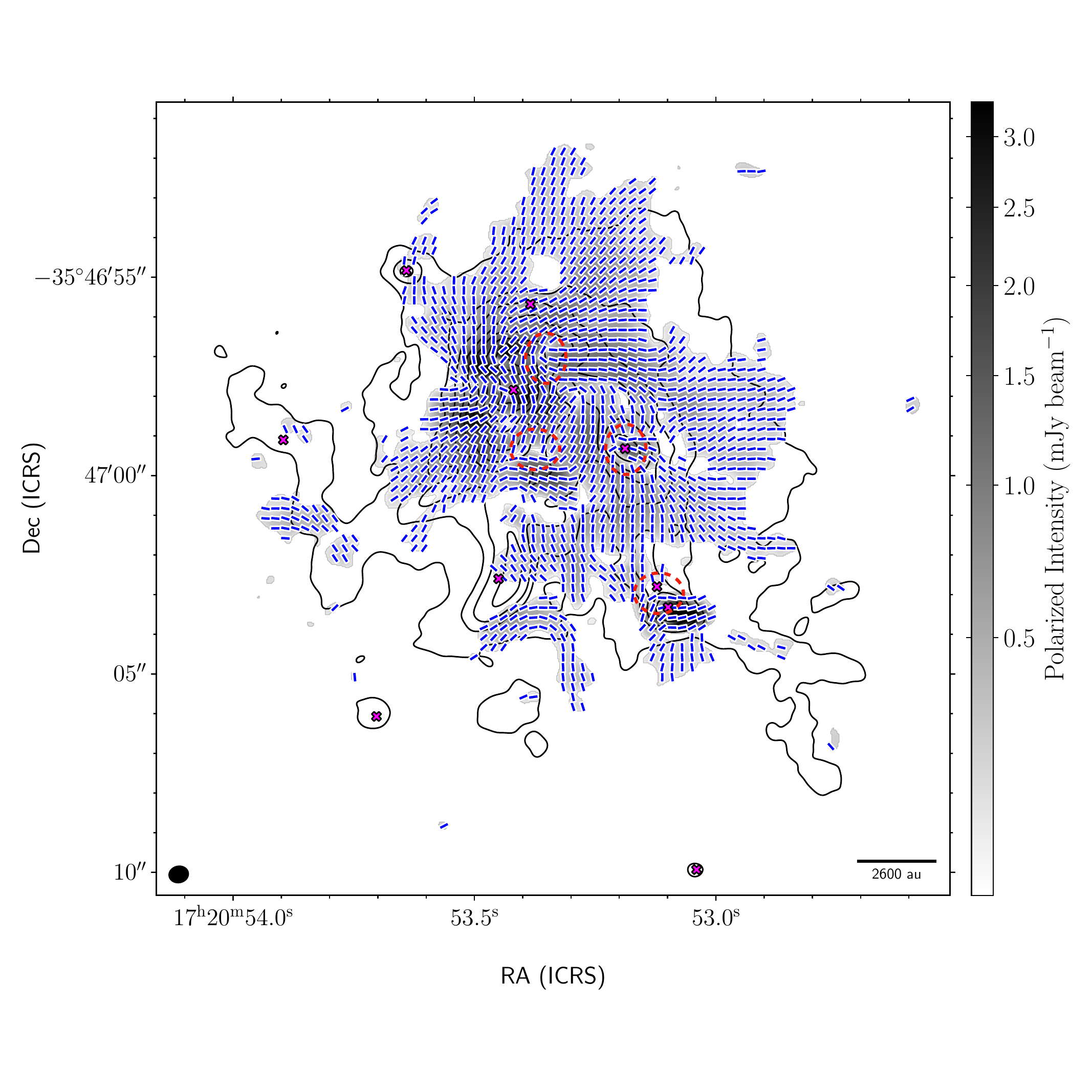}\\
\smallskip
\caption{
The figure illustrates the magnetic field morphology in NGC6334I, represented by blue pseudo-vectors. These are derived from a 90$^{\circ}$ rotation of the linearly polarized dust emission at 250 GHz, assuming the grains are aligned with the magnetic field. Pixel-level debiasing has been applied to the polarized emission at the 3$\sigma$ level. The resolution of the observation is indicated by the black ellipse in the lower left corner, representing the synthesized beam, while the scale bar in the lower right corner corresponds to an angular scale of  $2^{\prime\prime}$, or 2600 au. Positions of key sources are marked with magenta crosses and regions where the field appears to turn by $90^{\circ}$ are indicate by red segmented ellipses.
\label{fig:NGC6334I_POL}
}
\bigskip
\end{figure*}

\begin{figure*} 
\includegraphics[width=0.95\hsize]{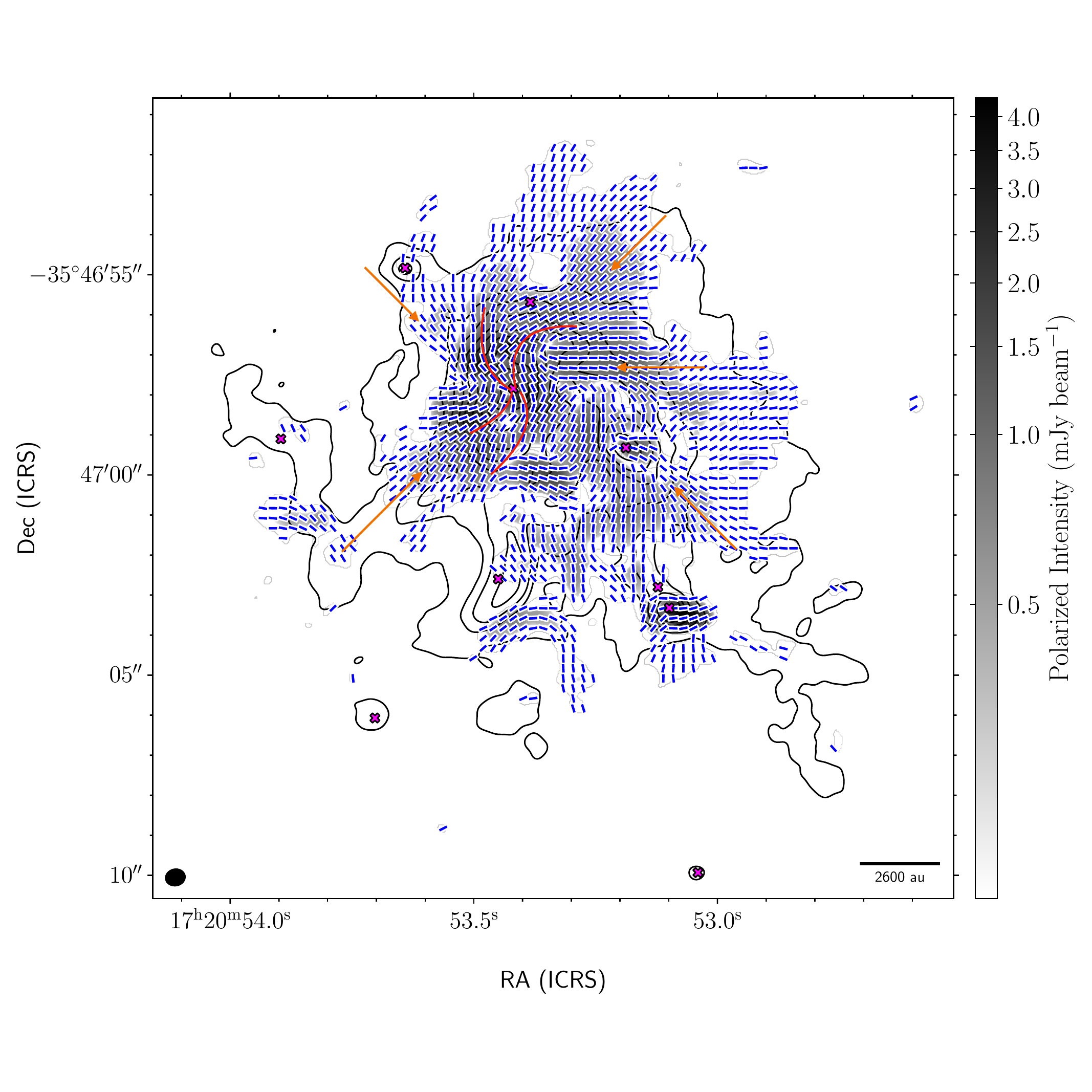}\\
\smallskip
\caption{
Same as Figure \ref{fig:NGC6334I_POL}, but here we indicate possible preferred directions of the field projected onto the plane of the sky. The orange arrow indicate radial direction towards the center of mass, the MM1-proto cluster. The red lines indicate possible spiral arms at the center of the MM1-proto cluster in NGC6334I.
\label{fig:NGC6334I_POL_2}
}
\bigskip
\end{figure*}

Figure \ref{fig:NGC6334I_POL} displays the magnetic field morphology on the plane of the sky, derived from polarized dust emission under the assumption of grain alignment by magnetic fields \citep{Lazarian2007}, which assumes a 90$^{\circ}$ rotation in the linear polarization position angle. While alternative alignment mechanisms such as radiative alignment \citep{Tazaki2017} and self-scattering \citep{Kataoka2016} have been proposed, the scales and physical conditions in massive star-forming regions (like the radiation anisotropy and large grain sizes), make them less favored. Consequently, magnetic alignment emerges as the most probable process driving the polarized emission detected towards NGC6334I. The polarized emission encompasses most of the primary sources in NGC6334I, including MM1, MM2, MM4, MM6, MM7, MM9, and CM2 (a strong water maser source with an ambiguous origin). Notably, the cometary UC{\HII} region MM3 lacks significant polarized dust emission, thus leaving its magnetic field untraced \citep[refer to Figure 1 in ][ for the extent of the MM3 UC{\HII}]{Sadaghiani2020}. The field morphology, albeit intricate, reveals consistent patterns across certain directions. For instance, the field demonstrates an evolution from North-West to South-East, showing discernible changes over the fragmented MM1 core, possibly a radial morphology towards the MM1 center of mass (see Figure \ref{fig:NGC6334I_POL_2}). Similarly, indications of radial configurations in the field lines can also be seen around the MM2 and MM4/MM6 cores. Radial morphologies suggests that gravity dominates and drags the field towards the center of mass \citep{Tang2009,Girart2013,Koch2018,Cortes2016,Cortes2019,Koch2022}. Noteworthy discontinuities, such as roughly $90^{\circ}$ shifts in field direction between adjacent pseudo-vectors, appear at the fringes of MM1 over the MM2 core and between MM6 and MM1. Despite these discrepancies, the field predominantly appears coherent on scales of $5^{\prime\prime}$ -- $10^{\prime\prime}$ (equivalent to 6500 to 13000 au, as visualized in Figures \ref{fig:NGC6334I_I} and \ref{fig:NGC6334I_POL}).  Spiral field patterns, like those observed in G327 and  IRAS 18089 \citep{Beuther2020,Sanhueza2021}, where spiral attributes are conspicuous and where rotation and infall have been deemed dynamically significant, maybe present at the inner regions of the MM1 proto-cluster  (see Figure \ref{fig:NGC6334I_POL_2}). However, clear rotation signatures are not conclusive from our data. Nonetheless, small scale spiral patterns have been observed in  other high mass  sources. \citet{Burns2023} observed spiral signatures on the also out-bursting source  G358.93-0.03-MM1 from high spatial resolution (50 to 900 au), 6.7 GHz methanol maser emission. 
Based on these results, it is plausible that, in NGC6334I, the indications of spiral morphologies seen in the field are retain in the gas, but not resolved due to the limitations of the ALMA configuration used here.

\subsection{Molecular Line Emission}
\label{sse:mol_emission}

The spectral setup used by the MagMar project allow the detection of molecular line emission from a number of tracers usually abundant in high mass star forming regions. From our data, we used CS$(J=5 \rightarrow 4)$ and C$^{33}$S$(J=5 \rightarrow 4)$ to study outflow and non-thermal motions. 
From the data obtained by \citet{Liu2023b}, we extracted the $^{12}$CO$(J=2 \rightarrow 1)$ emission which we used to study the outflows present in NGC6334I.

\setlength{\tabcolsep}{2pt}
\begin{deluxetable*}{cccccccc}
\tablecolumns{8}
\tablewidth{0pt}
\tabletypesize{\scriptsize}
\tablecaption{ Flux Variability \label{table:fluxVar}}
\tablehead{
\colhead{Stokes} &
\colhead{Max$_{\mathrm{s1}}$} &
\colhead{Mean$_{\mathrm{s1}}$} &
\colhead{$\sigma_{\mathrm{s1}}$} &
\colhead{Max$_{\mathrm{s2}}$} &
\colhead{Mean$_{\mathrm{s2}}$} &
\colhead{$\sigma_{\mathrm{s2}}$} &
\colhead{variability} \\
\colhead{}     &
\colhead{($\mathrm{mJy\,beam}^{-1}$)} &
\colhead{($\mathrm{mJy\,beam}^{-1}$)} &
\colhead{($\mathrm{mJy\,beam}^{-1}$)} &
\colhead{($\mathrm{mJy\,beam}^{-1}$)} &
\colhead{($\mathrm{mJy\,beam}^{-1}$)} &
\colhead{($\mathrm{mJy\,beam}^{-1}$)} &
\colhead{(\%)}
}
\startdata
I & 1330 & 45 & 0.988 & 1331 & 42 & 1.0 & 0.1\\
Q & 3.91 & 0.028 & 0.067 & 4.3 & -0.007 & 0.073 & 8\\
U & 3.97 & 0.075 & 0.083 & 4.3 & 0.088 & 0.068 & 8\\
V & -1.06 & -0.003 & 0.083 & 2.8 & 0.06 & 0.100 & -
\enddata
\vspace{0.1cm}
\tablecomments{The flux density variability for all Stokes continuum images is listed listed here. The values where extracted from the images considering the 1/3 FWHM for the maximum and the mean while we used a region devoid of emission for $\sigma$. The subscript s1 and s2 indicates session 1 and session 2. Note, Stokes $Q$, $U$, and $V$ can be negative and thus, the maximum is not necessary a positive value.  We do not list the variability in Stokes V due to the large ALMA uncertainties with circular polarization. }
\end{deluxetable*}

\begin{figure*}[h!]
\includegraphics[width=0.8\hsize]{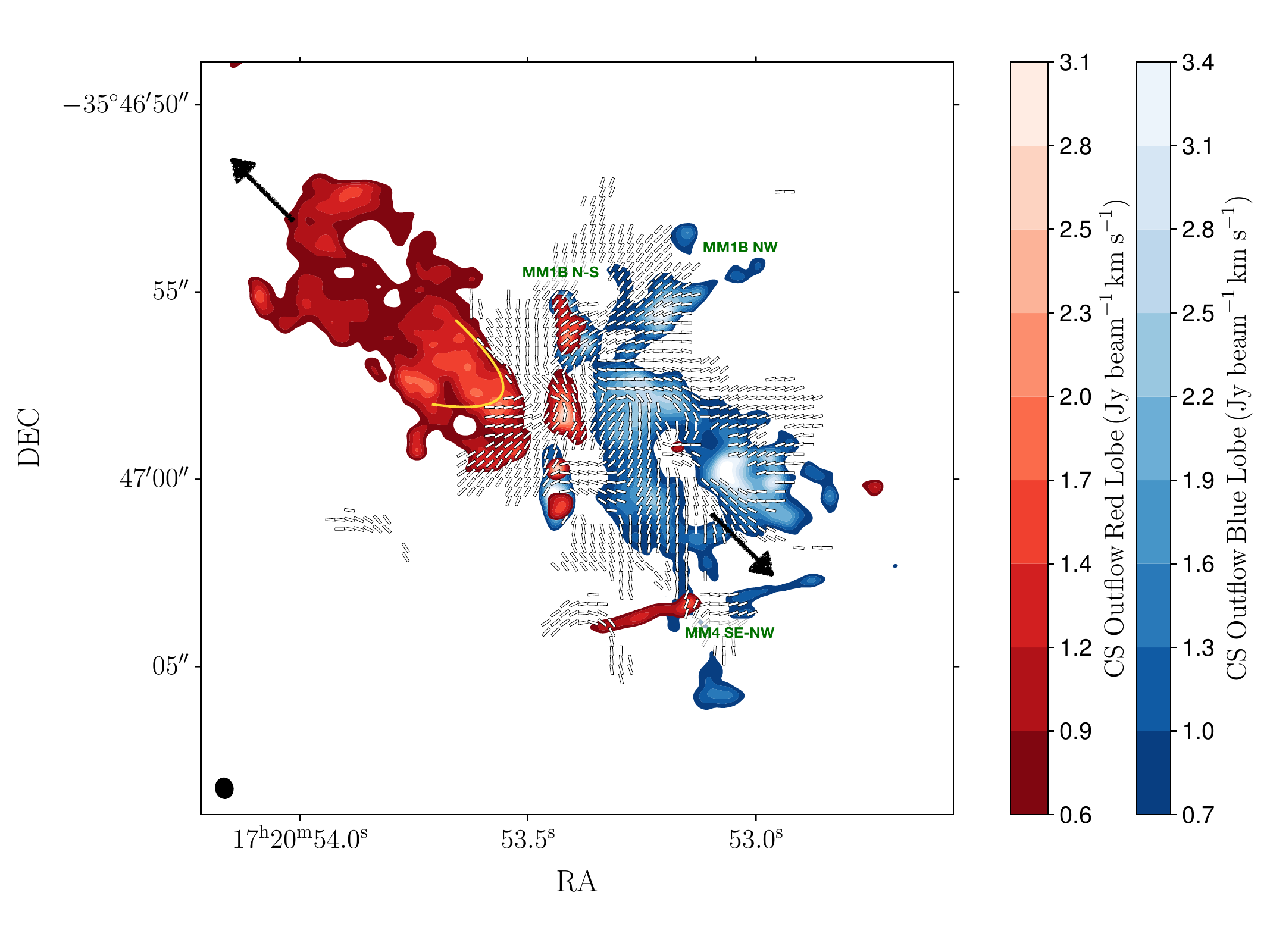}\\
\includegraphics[width=0.8\hsize]{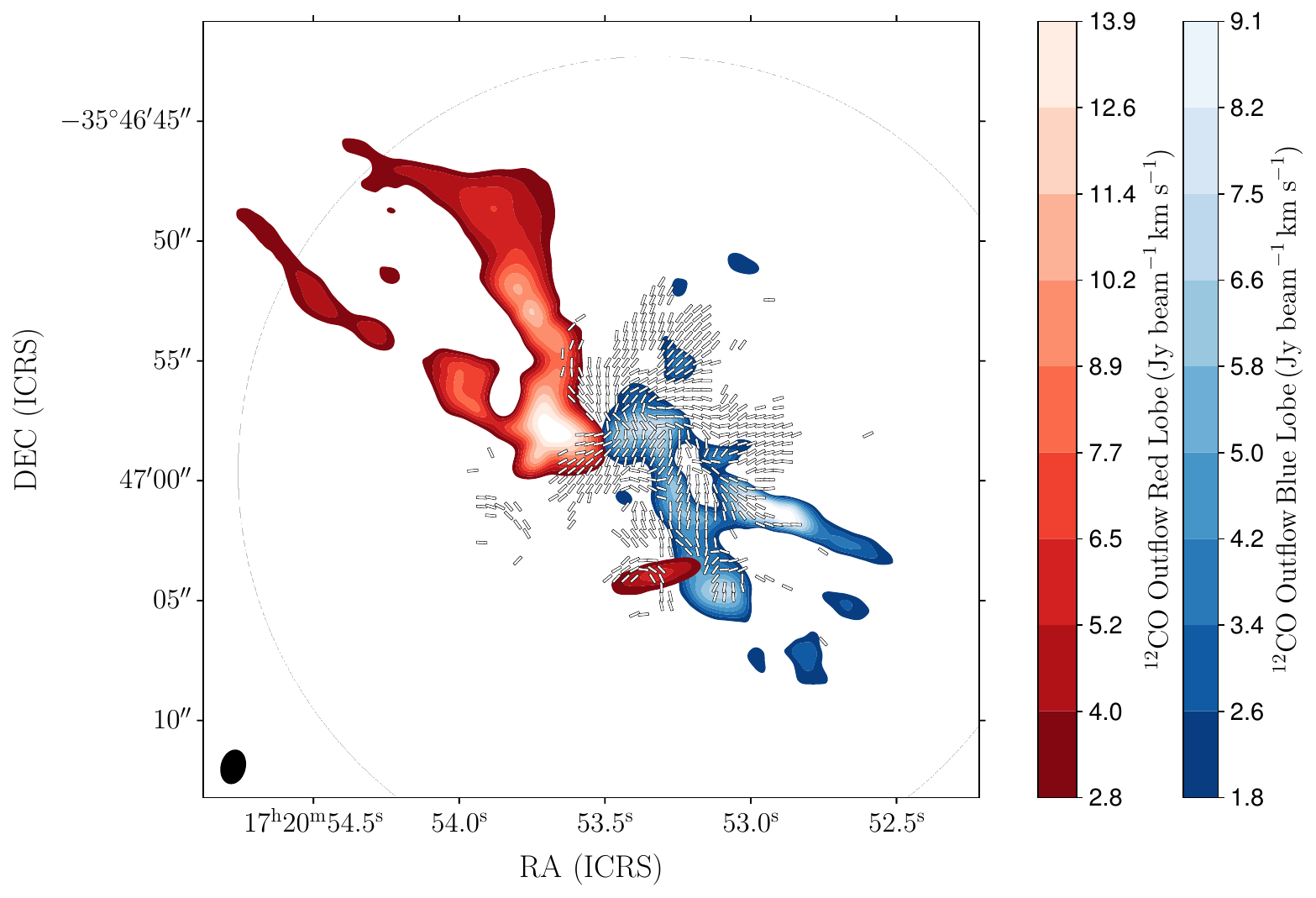}
\smallskip
\caption{
{\bf\em Top:} CS($J=5 \rightarrow 4$) emission from NGC6334I's outflows  with the magnetic field overlaid. Blue and red lobes are derived from velocities -40 to -12 \kms\ and -4 to 12 \kms, respectively. The outflow cavity is outlined in yellow, with individual flows labeled per \citet{Brogan2018} and the primary NE-SW flow indicated by arrows.
{\bf\em Bottom:} CO($J=2 \rightarrow 1$) emission tracing NGC6334I's outflows with with the magnetic field overlaid. Blue and red lobes span from -45 to -25 \kms\ and -4 to 11 \kms, respectively. The CO emission also reveals the cavity, with the magnetic field contouring the outflow's edges.
\label{fig:outflow}
}
\bigskip
\end{figure*}

\subsubsection{The $\mathrm{CS}(J=5 \rightarrow 4)$ Emission}
\label{sse:cs}
Figure \ref{fig:outflow} displays a moment zero map of the CS$(J=5 \rightarrow 4)$ emission towards NGC6334I, where the data have been categorized into redshifted and blueshifted velocity components to show outflow emission. This categorization was done by considering a systemic velocity of -7.56 {\kms}, as determined by a Gaussian fit to the C$^{33}$S line (refer to section \ref{sse:c33s}), with velocity intervals of -40 to -12 {\kms} for the blueshifted and -4 to 12 {\kms} for the redshifted components.
Outflow emission is clearly seen in the CS moment zero map, where
our data align with the morphological findings of \citet{Brogan2018}. Following their nomenclature, our map  reproduces the prominent NE-SW, the MM1B N-S, the MM1B NW, and the MM4 SE-NW outflow emissions as indicated in Figure \ref{fig:outflow}.
From the four outflows detected towards NGC6334I, the NE-SW outflow appears to be dominant. Although its exact origin has not been conclusively determined, it appears to emerge from the MM1 proto-cluster which is also the likely origin of the smaller
MM1B N-S and MM1B NW outflows.
From the map, the NE-SW outflow red-lobe appears to be carving a cavity in the dust as shown in Figure \ref{fig:outflow}. The edges of this cavity seem to be traced by magnetic fields as we will show later (see section \ref{se:discussion}).
To study the CS emission 
in more detailed, we modeled the data using the MADCUBA package which fits spectroscopy models to the data using a complete radiative transfer formalism under a Local Thermodynamic Equilibrium (LTE) assumptions \citep{Martin2019}. To this effect,
we extracted a spectrum from the central region of NGC6334I containing the CS line and a number of other molecular transitions
(see Table \ref{table:setup} for a description of the spectral windows containing the CS line). 
This spectrum was modeled with the MADCUBA package where we found that the CS line would be optically thick ($\tau = 6$) assuming an excitation temperature  $T_\mathrm{ex} = 260 \pm 20$ K 
derived with available transitions of $\rm CH_3OH$, $\rm ^{13}CH_3OH$, and $\rm C_2H_5OH$ in the observed spectral windows, and based on the C$^{33}$S fit and assuming a 
$^{32}$S/$^{33}$S=102 taken from \citep{Yu2020}, since the CS profile is heavily affected by self-absorption.

\begin{figure*} 
\includegraphics[width=0.5\hsize]{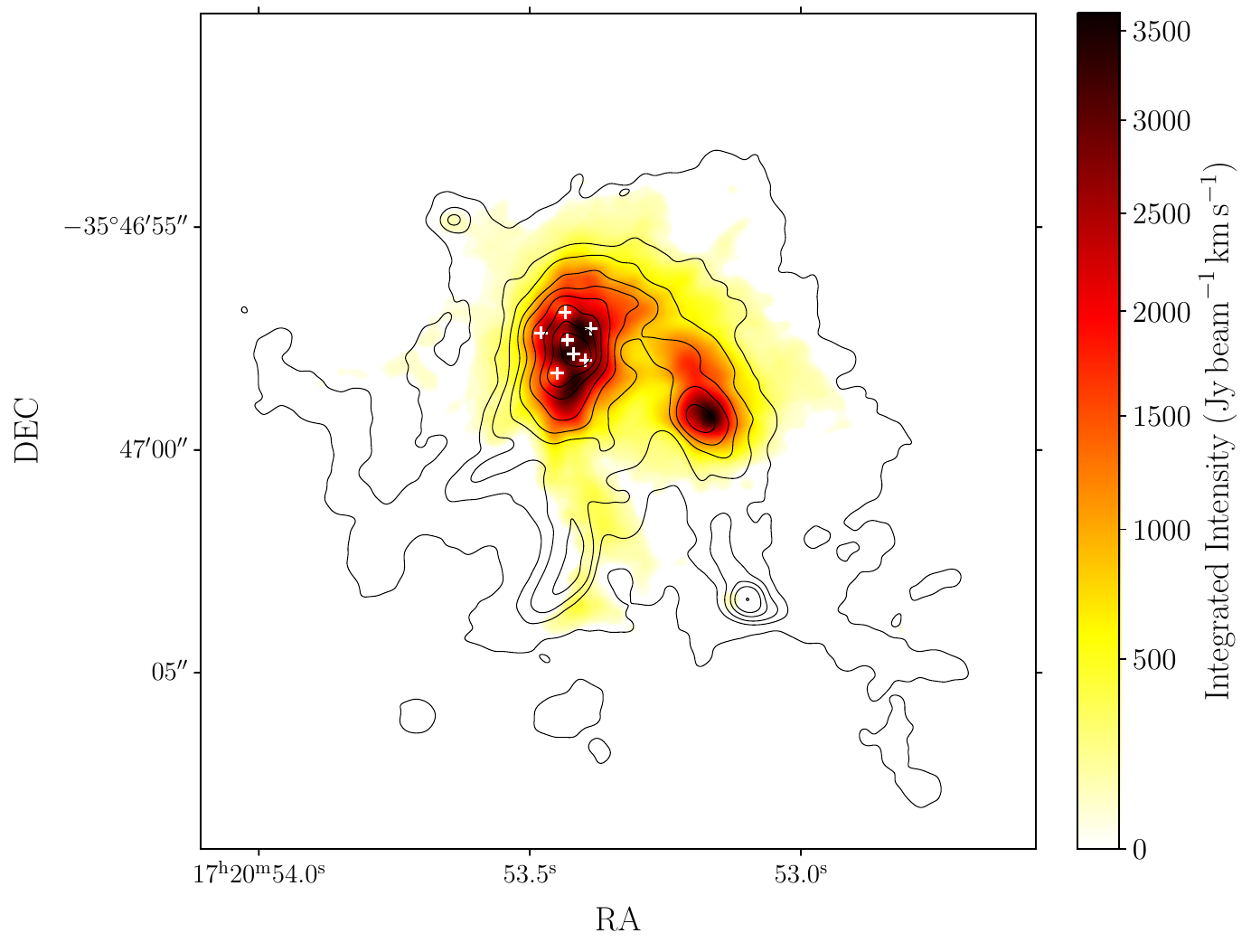}
\includegraphics[width=0.49\hsize]{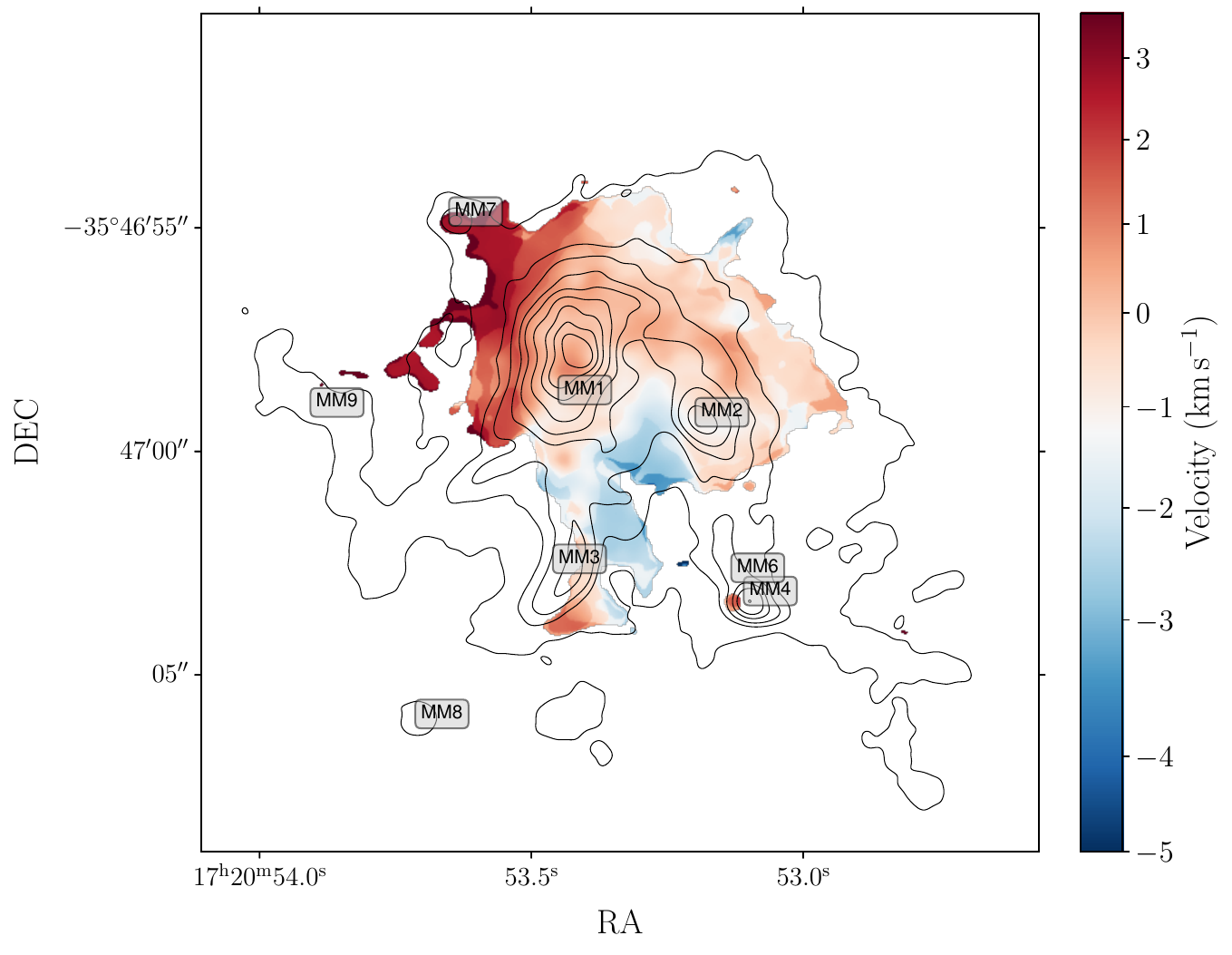}
\smallskip
\caption{The figure shows to the {\bf left} the C$^{33}$S moment 0 map in color scale with superposed contours from dust emission starting at the $3\sigma$ level (see Figure \ref{fig:NGC6334I_I}). The white crosses correspond to the MM1 proto-cluster condensations resolved by \citet{Brogan2016A}. To the {\bf right}, we show the moment 1 map in color scale, also with dust continuum contours superposed, as $\Delta V_{i,j} = V_{i,j} - V_{\mathrm{lsr}}$, where $V_{\mathrm{lsr}}= -7.56$ is the velocity center of the C$^{33}$S line as determined by a Gaussian fit. 
\label{fig:c33s_moms}
}
\bigskip
\end{figure*}

\subsubsection{The $\mathrm{C}^{33}\mathrm{S}(J=5 \rightarrow 4$) Emission}
\label{sse:c33s}
\begin{figure*} 
\includegraphics[width=0.99\hsize]{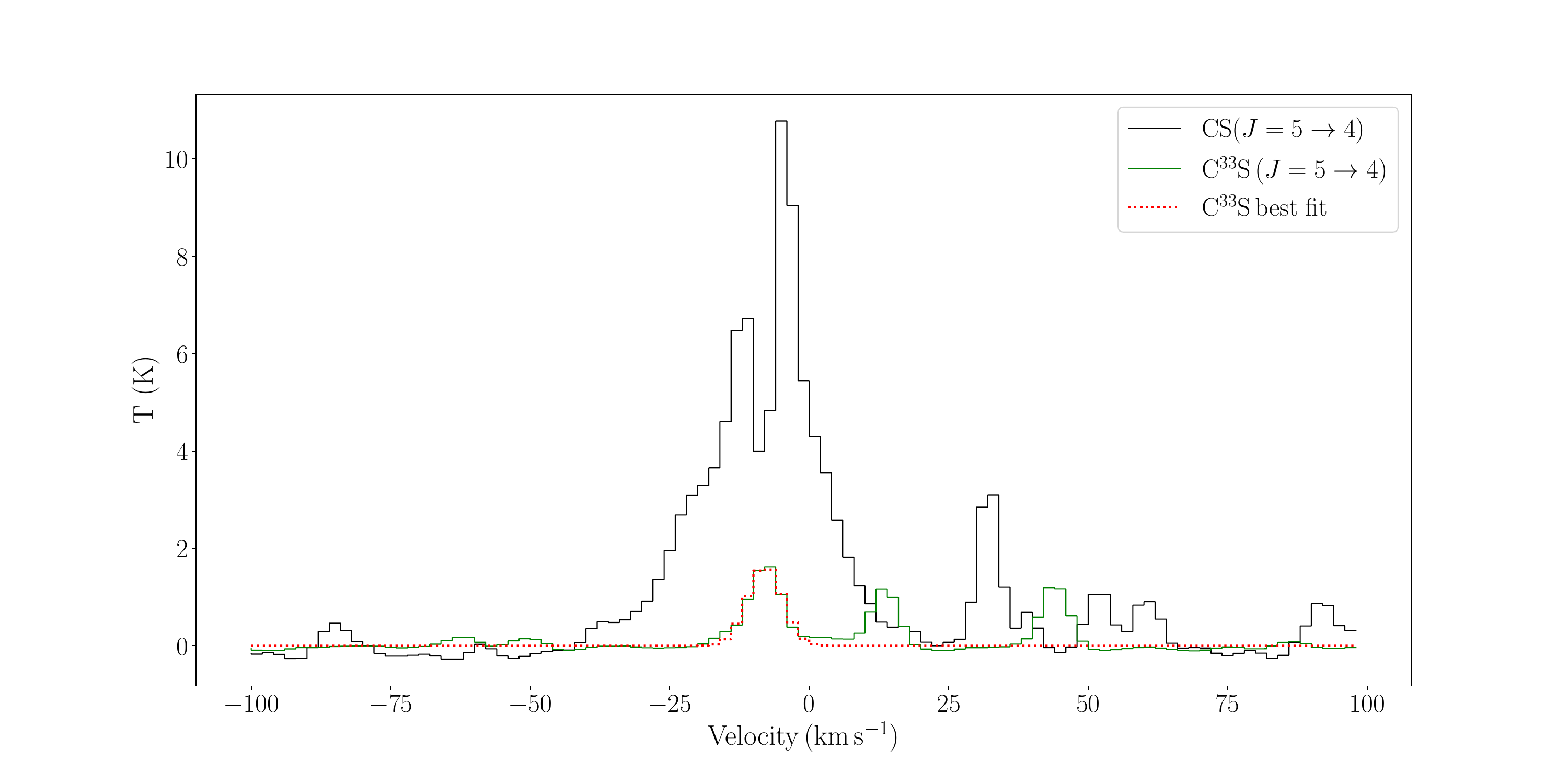}
\smallskip
\caption{The Figure shows spectra from  CS$(J=5 \rightarrow 4)$ (black), C$^{33}$S$(J=5 \rightarrow 4)$ emission (green), and its best Gaussian fit (red dotted line) extracted from an area covering the outflow emission seen from CS moment 0 map.  The other two spectral features seen in the C$^{33}$S spectrum correspond to Methyl Formate emission \citep[similar to what was seen in NGC6634I(N), see ][]{Cortes2021b}.
\label{fig:outflowC33S}
}
\bigskip
\end{figure*}

The C$^{33}$S($J=5 \rightarrow 4$) emission appears to be compact and confined to the brightest dust emission contours, as shown in Figure \ref{fig:c33s_moms}. The peaks in the integrated intensity map coincide with the dust emission peaks while most
of the C$^{33}$S is confined within the $15\sigma$ contour from the dust emission. While the MM1 proto-cluster is not resolved in the dust emission, three distinct condensations are observable in C$^{33}$S from North to South and around the dust peak, which are loosely associated with the condensations resolved by the higher resolution ALMA data \citep{Brogan2016A} as indicated by the white crosses in Figure \ref{fig:c33s_moms}. Toward MM2, the C$^{33}$S peak emission aligns well with the dust maxima, whereas the emission levels appear marginal toward MM4 and MM7, and remain undetected toward MM9. In Figure \ref{fig:c33s_moms}, the moment 1 map is also presented, relative to V$_{\mathrm{lsr}} = -7.56, \kms$, obtained through Gaussian fitting of the C$^{33}$S spectrum from the 1/3 FWHM of the primary beam. From this map, a velocity 
gradient is seen from North-East to South-West which is consistent with that obtained by \citet{Liu2023b} using the OCS molecular emission, though our coverage does not match the extent of the ALMA mosaic in \citet{Liu2023b}. Although the velocity gradient might have some outflow contamination, the spectra shown in Figure \ref{fig:c33s_moms} suggests it should be minimal. This is because the C$^{33}$S line profile is almost completely Gaussian with very small deviations around the -20 {\kms} velocity channel.
Thanks to the lack of absorption features, we could fit the C$^{33}$S spectrum  with the same temperature assumption done for CS above, which suggests that the emission is optically thin (tau=0.3) which allows to use the moment 2 map as the velocity dispersion along the line of sight (see Figure \ref{fig:outflowC33S}).

\subsubsection{The $^{12}\mathrm{CO}(J=2 \rightarrow 1$) Emission}
\label{sse:co}

\begin{figure*} 
\includegraphics[width=0.95\hsize]{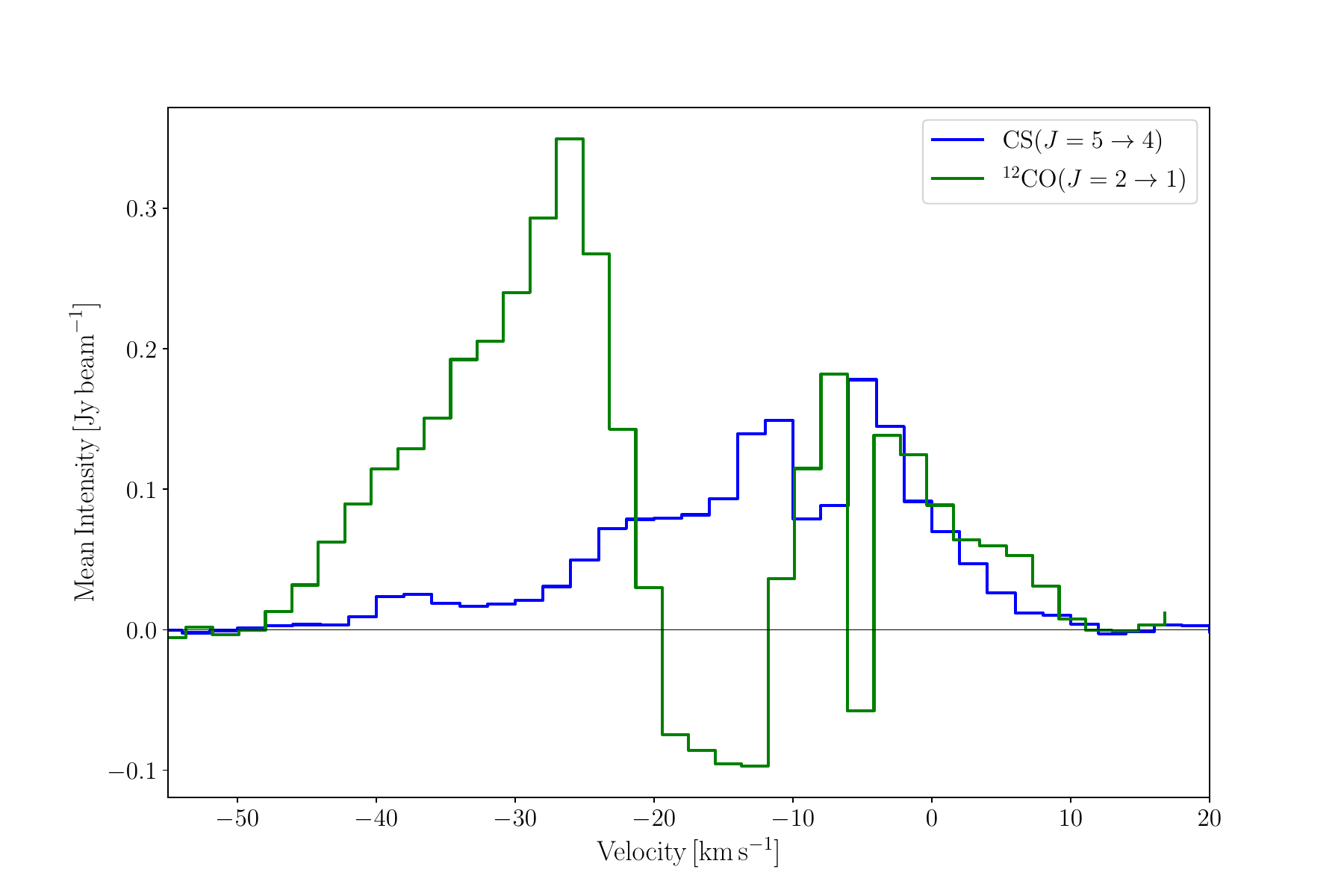}
\smallskip
\caption{The Figure shows spectra from the $^{12}$CO$(J=2 \rightarrow 1)$ and CS$(J=5 \rightarrow 4)$ emission extracted as the mean within the 1/3 FWHM of our maps, corresponding to 0.05 pc. The $^{12}$CO spectrum was binned to match the velocity resolution of  the CS spectrum at 2 \kms\,. 
\label{fig:cs_co_Outflow}
}
\bigskip
\end{figure*}

We used the  ALMA data obtained by \citet{Liu2023b} to image the outflow emission from NGC6334I (see Table \ref{table:setup} for a description of their data). To image the outflow emission we used a velocity interval from -45 to -25 {\kms} for the blue lobe and -4 to 11 {\kms} for the red lobe (refer to Figure \ref{fig:outflow}, which displays the combined data from both ALMA configurations).  From these data, only the NE to SW outflow is clearly discernible, whereas the N-S outflow in the line of sight remains undetected in $^{12}$CO and some traces of the MM1B NW outflows are seen in the blue-lobe. Additionally, only the red-lobe of the MM4 outflow is observable in $^{12}$CO. 
As with the CS emission, the red-lobe of the CO outflow appears to be carving a cavity in the dust where the magnetic field traces the edges of the cavity (see section \ref{se:discussion}).
Although this data set includes configuration C41, the most compact ALMA configuration with a Maximum Recoverable Scale (MRS) of $\sim 9^{\prime \prime}$ (0.06 pc at 1300 pc, the distance to NGC6334I), there appears to be significant extended emission from $^{12}$CO that the interferometer is resolving out. 
Figure \ref{fig:cs_co_Outflow} shows CO and CS spectra superposed to each other. Although the outflow emission is seen in both molecular profile line-wings, the comparison suggests, as previously mentioned, that significant emission from CO is missing. This is likely either due to filtering effects or self-absorption (see section \ref{sse:co}). Therefore, we omit using CO emission for the outflow analysis here.

\section{DISCUSSION}\label{se:discussion}

At first glance, the magnetic field in NGC6334I displays a complex morphology 
which is difficult to associate with a projection from a simple shape, such as the ``hourglass'' proposed by models \citep{Ciolek1994} and observed in numerous star-forming regions \citep{Girart2006,Hull2014,Li2015,Beltran2019,Cortes2021b}. Although there are indications of a radial pattern at the outskirts of the clump and  in the direction of the MM1 proto-cluster,  there are also hints of some sort of spiral arms in the field close to the central region of the MM1 proto-cluster (see Figure \ref{fig:NGC6334I_POL_2}). The radial pattern can be explained by a dominant gravitational pull, but for the spirals, we do not find conclusive evidence of rotation or velocity gradients that can explain these features in the field morphology at these scales (see Figure \ref{fig:c33s_moms}). 
Nonetheless, small scale spiral patterns cannot be ruled out given previous finding in other similar high mass out-bursting sources \citep[see ][]{Burns2023}. 
Additional complexity in the field morphology arises from  sudden 90$^{\circ}$ shifts in the field lines. These turns in the field morphology can be seen around the MM2, MM6 cores, and also at the interface of the MM1 proto-cluster as indicated by the red ellipses in Figure \ref{fig:NGC6334I_POL}. These sharp shifts in the field morphology are challenging to elucidate; they could result from projection effects or perturbations in the field by the gas due to turbulence or gravity. 
One possibility is that the four outflows detected in this region might be perturbing the magnetic field structure.
For instance, the most prominent bipolar outflow, seen in both CO and CS emission red lobe, appears to carve a cavity in the dust where the magnetic field seems to follow the edges of the cavity as indicated in Figure \ref{fig:outflow}. Similar situations have been seen in low mass star forming regions such Serpens SMM1-b where the outflow emission was also found to trace the edges at the base of the outflow cavity \citep{Hull2017b}.  
We also note that the field show some degree of alignment with the MM1B NW and the MM4 SE-NW outflows. However, whether this is a projection effect or real alignment cannot be concluded from these data.
As understanding projection effects demands extensive numerical modeling, we instead examine whether the energy balance in NGC6334I could explain the perturbations seen in the field.

\subsection{The Energy in the Outflows}
\label{sse:outpar}
To explore the quantitative effect of the outflow in the magnetic field, we attempt here to compute the outflow parameters in order to estimate their total energy.
To derive the energy in the  outflows, we used the CS emission rather than the CO emission. This is because the CO emission shows, what appears to be, significant filtering by the interferometer and possibly extensive self-absorption which will introduce larger errors in the estimations. Also, the CS line appears to better trace the outflows at the core scales mostly resolving all four outflows while the CO does not clearly show all of the outflow emission detected in NGC6334I (see Figure \ref{fig:outflow}). However, it should be noted that because the interferometer does not recover all of the CS emission and because of the uncertainties in the abundance ratio between CS and H$_{2}$, the outflow parameters derived here may have a large uncertainty.
To estimate the outflow parameters, we extracted spectra from the CS  velocity cube covering the extent of the CS moment 0 map
(see Figure \ref{fig:outflowC33S}). Determining the outflow parameters require separating the emission coming from the gas at the source systemic velocity and the actual emission from the outflow. It also requires estimating the optical depth in the line-wings which is usually done by computing the line ratio with an isotopologue that has sufficient overlap in the line-wings covering the outflow emission. Unfortunately, the C$^{33}$S emission detected towards NGC6334I is compact enough that the overlap with the CS at velocities associated with the outflow is minimal (see Figure \ref{fig:outflowC33S}). 
Thus, to estimate the CS column density corresponding to the outflow emission, we assume that this emission is optically thin in the channels covering the CS line-wings.
The line-wings of the CS line cover the interval from  -40 to -12 \kms\, for the blue lobe and -2 to 20 \kms\, for the red lobe where the systemic velocity of the source is determined to be -7.6 {\kms} as obtained from the Gaussian fit to the C$^{33}$S line (see Figure \ref{fig:outflowC33S}). This interval covers all of the emission in the line-wings over 3$\sigma$, with 
$\sigma = 0.2 $ K estimated from 25 line free channels.
Note, we are not aiming to characterize independently the four outflows in this region, but rather estimate their bulk energy  and its effect on the magnetic field.

By assuming that the emission is optically thin, we estimate the CS column density per velocity channel as \citep[see Equation 16, ][]{Zhang2016},

\begin{equation}
N = \left( \frac{8\pi k \nu_{ul}^{2}}{h c^{3} A_{ul} g_{u}} \right)  \frac{T_{R}(\nu)}{f} Q_{\mathrm{rot}}(T_{\mathrm{ex}}) \delta v e^{\frac{E_{u}}{k T_{\mathrm{ex}}}} 
\end{equation}

where $k$ is the Boltzmann constant, $h$ is the Planck constant, $c$ is the speed of light in the vacuum, $\nu_{ul}=244.9355565$ GHz is the rest frequency of the $J=5 \rightarrow 4$ CS rotational transition, $A_{ul} = 2.981 \times 10^{-4}$\, s$^{-1}$ is the Einstein spontaneous emission coefficient, $g_{u} = 11$ is the statistical weight, $E_{u} = 35.3$ K is the energy level for the upper state, $Q_{\mathrm{rot}}$ is the partition function, which for linear molecules is well approximated by $Q_{\mathrm{rot}}=
\frac{k T_{\mathrm{ex}}}{h B_{0}}$, where $B_{0} = 24.5$ GHz is the rotational constant for the CS molecule \citep{Bustreel1979}, $\delta v = 2\, \kms\,$ is the channel width, and $f$ is the beam filling factor assumed
to be 1. 
As indicated in Section \ref{sse:cs}, the excitation temperature, T$_{ex} = 260$ K, is estimated from the modeling done with MADCUBA on symmetric rotor molecules.
The mass of the outflow is calculated for every velocity channel as:

\begin{equation}
M = \mu_{\mathrm{H_{2}}}m_{H}AN_{\mathrm{H_{2}}}
\end{equation}

where $\mu_{\mathrm{H_{2}}} = 2.8$ corresponds to the mean molecular weight \citep{Feddersen2020}, $m_{\mathrm{H}}$ is the Hydrogen mass, and $A$ is the area subtended by the region identified with the outflows (same used to extract the spectra from Figure \ref{fig:outflowC33S}), and $N_{\mathrm{H_{2}}} = N/X^{\mathrm{CS}}_{\mathrm{H_{2}}}$ is the total column density of the molecular Hydrogen where $X^{\mathrm{CS}}_{\mathrm{H_{2}}} = 1.2\times10^{-7}$ is the relative abundance of CS respect to the H$_{2}$ in NGC6334I as determined from Herschel mapping \citep{Zernickel2012}. The momentum and kinetic energy per velocity channel is estimated as $P_{i} = M_{i}v_{i}$ and
$E_{i} =M_{i}v^{2}_{i}/2$ where $i$ indicates the velocity channel used for each quantity \citep[in this case $M_{i}$ is the mass spectrum, see ][]{Feddersen2020}. The total mass, momentum, and energy is obtained by adding all velocity channels per lobe. 
The values for the mass energy and momentum are listed in Table \ref{table:outflow}. 
The assumption that the CS emission is optically thin the line-wings appears to be good enough to produce under-estimates of
the total energy in the outflow when compared to single dish results. \citet{McCutcheon2000} estimated the kinetic energy on each outflow lobe to be $\sim 1.4 \times 10^{47}$ ergs from their CO and CS millimeter single dish mapping while our estimates are an order of magnitude lower (see Table \ref{table:outflow}). 
Although,  the mass, momentum, and energy are consistent with outflow properties from surveys of high mass protostellar outflows \citep{Zhang2001,Zhang2005}, it should be noted that we are missing emission that might be present at the systemic velocity as well as the emission filtered out by the interferometer. Thus, the values derived here appear to be a lower bound for the outflow mass, momentum, and energy at these scales in NGC6334I.

\setlength{\tabcolsep}{2pt}
\begin{deluxetable*}{ccccc}
\tablecolumns{5}
\tablewidth{0pt}
\tabletypesize{\scriptsize}
\tablecaption{ Outflow Parameters\label{table:outflow}}
\tablehead{
\colhead{Source} &
\colhead{Lobe} &
\colhead{$M$} &
\colhead{$P$} &
\colhead{$E$} \\
\colhead{}     &
\colhead{}     &
\colhead{(\Msun)} &
\colhead{(\Msun\, \kms\,)}     &
\colhead{($10^{46}$ ergs)}     
}
\startdata
NGC6334I & Blue & 0.3 & 6.4 & 0.24  \\
NGC6334I & Red & 0.6 & 4.9 & 0.11
\enddata
\tablecomments{The Outflow parameters determined from the CS emission are presented here. The red lobe was obtained by considering a velocity range between -4 to 12 {\kms} while the blue lobe was obtained from a range between -40 to -12 {\kms}.
}
\end{deluxetable*}

\subsection{The Magnetic Field Strength Map}
\label{sse:bpos}

To estimate the magnetic field strength on the plane of the sky, B$_{\mathrm{pos}}$, we employed the Davis, Chandrasekhar, and Fermi technique \citep[DCF,][]{Davis1951,Chandrasekhar1953} as outlined by \citet{Crutcher2004}. In units of $\mu$G, this can be expressed as:

\begin{equation}
\label{eq:Bpos}
\mathrm{B}_{\mathrm{pos}} = 9.3 \sqrt{n_{\mathrm{H_{2}}}}\Delta\mathrm{V}/\delta\phi.
\end{equation}
\bigskip

where $n_{\mathrm{H_{2}}}$ is the number density in cm$^{-3}$, $\Delta\mathrm{V}$ is the molecular linewidth from an optically thin species in \kms\, and $\delta\phi$ is the dispersion in the magnetic field lines in degrees.
Even though the DCF method has been revisited over time to explore its limitations and applicability \citep[see reviews by ][]{Liu2021, Myers2024}, we aimed to mitigate biases in this work by simplifying our analysis and minimizing assumptions. Consequently, we employed the ``standard'' DCF method, incorporating only the 1/2 correction proposed by numerical simulations \citep[see][for a comprehensive discussion]{Crutcher2004}. We will
explore the implication of this in section \ref{sse:caveats}.

Our objective here is to derive a map with estimations of the magnetic field strength onto the plane of the sky. To do that, we employ the 
number density model map, the velocity dispersion, $\Delta \mathrm{V}$, obtained from the C$^{33}$S moment 2 map (see
Figure \ref{fig:vdisp}), and a position angle dispersion map, $\delta \phi$. We use the C$^{33}$S  emission because is optically thin (see section \ref{sse:c33s}), and thus it provides a good account of the non-thermal motions of the warm and dense gas. Although as we inspected the data cube only a single component of the line was observed, it is difficult to rule out, or remove, the small outflow contribution to the C$^{33}$S  emission. Nonetheless and because the emission is compact with very small deviation from the model at the linewings (see section \ref{sse:c33s}), we assume that this contamination is minimal. Although it does not overlap completely with the polarized dust emission, the C$^{33}$S emission has sufficient coverage around the most relevant cores in NGC6334I to provide a good account of turbulence in this region. 

\begin{figure*}[h!] 
\includegraphics[width=0.95\hsize]{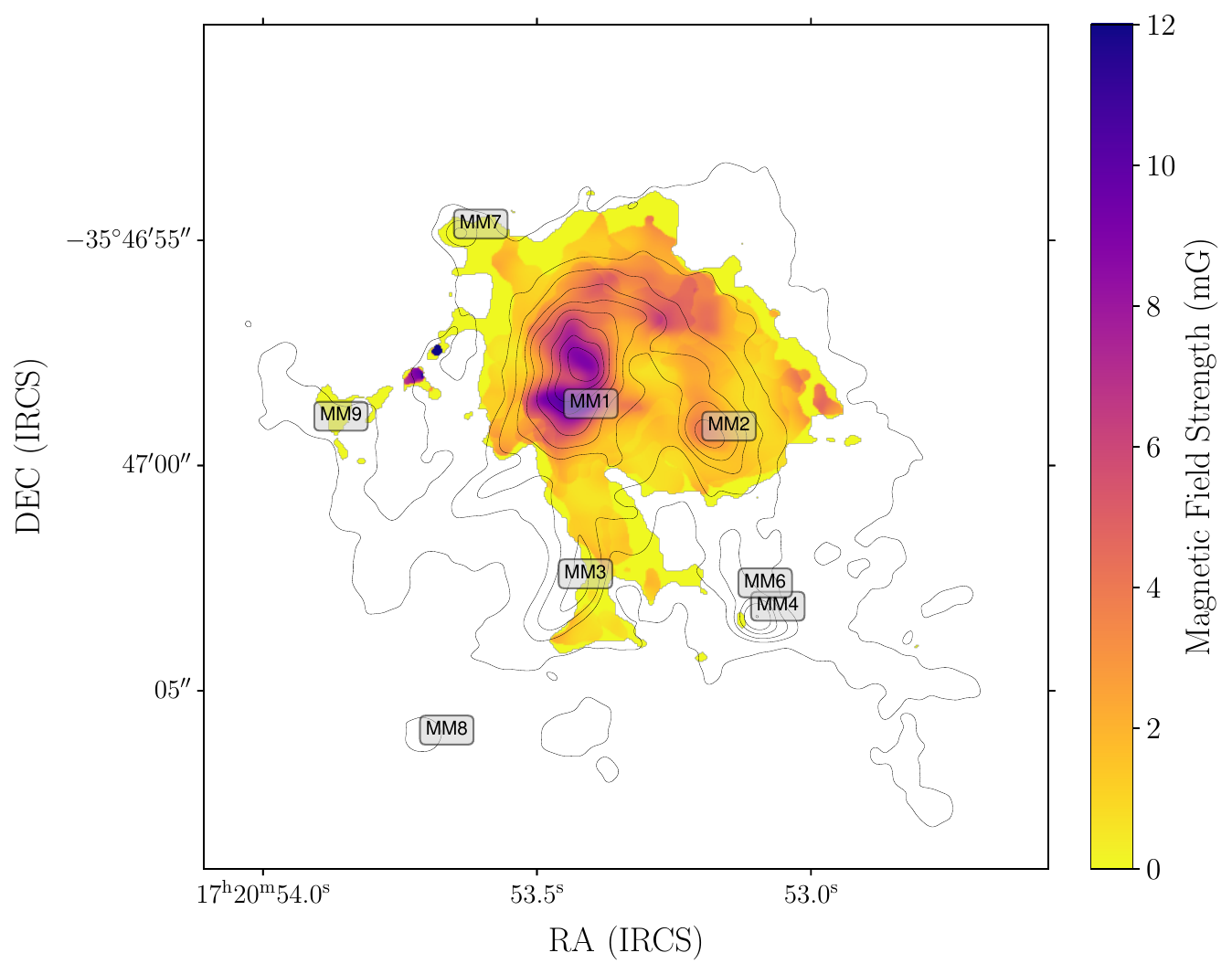}
\smallskip
\caption{The Figure shows the estimated field strength map onto the plane of sky determined by using the DCF method over a moving window of 1$^{\prime\prime}$.5 used to estimate the position angle dispersion by means of circular statistics. Superposed to the map we add the dust continuum emission as contours following Figure \ref{fig:NGC6334I_I}
and the source labels as previously shown in Figure \ref{fig:NGC6334I_I}.
\label{fig:bpos}
}
\bigskip
\end{figure*}

The dispersion in the magnetic field lines is obtained by taking the standard deviation over a moving window of $1^{\prime\prime}.5$ in size. This kernel is $\sim 4\times$ the beam size, which give us about 16 independent points per window to calculate the standard deviation using circular statistics under the assumption of a 5$^{\circ}$ error in the polarization position angle (see Appendix \ref{ap:std} for a discussion about the statistics of the moving window). The window size used here has a sufficient statistical
significance to produce a believable estimate for the angular dispersion, but also encloses the local fluctuations in the
field, which are seen to occur in angular scales of $\sim 1^{\prime\prime}$ by visual inspection. 
While our goal is to account for the best possible estimate of the magnetic field lines local dispersion, we also strive to prevent skewing our results in areas where the magnetic field 
appears uniform. To achieve this, we are cautious about selecting a larger moving window size, which could inadvertently include areas exhibiting significant fluctuations in the magnetic field's direction
(i.e. regions where the
field shows $\sim 90^{\circ}$ turns). Therefore, a value of $1^{\prime\prime}.5$ in size for the moving window appears to be a good compromise to achieve these objectives (Appendix \ref{ap:std} offers a more detailed discussion showing the effect of increasing the moving window size).
Note, it is entirely plausible that the sudden turns in $90^{\circ}$ are due to projection effects and not because of perturbation in the field lines.
If that were to be the case, the estimated dispersion from that regions would be artificially large, which will decrease the estimated field strength and thus the values around those regions should be considered as lower bounds for the magnetic field strength onto the plane of the sky. 
Thus and under these considerations, the B$_{\mathrm{pos}}$ map
estimate is shown in Figure \ref{fig:bpos}. The mean magnetic field strength estimated value  is $\sim 1.9$ mG with a peak of $\sim$ 11 mG towards the MM1 proto-cluster. Note, a strong outlier to the East of MM1 appears to be the result of small statistics at that point which we ignore. Although \citet{Li2015} estimated B$_{\mathrm{pos}} = 12$ mG at scales of 0.1 pc from the SMA data, our mean value is lower which can be explained because of the coarser SMA resolution that smooths the field morphology and also because of their distance assumption to NGC6334 (1.7 kpc) which yields and overestimation of the mean density. Our estimate is consistent with the magnetic field strengths reported in a more recent analysis of the SMA data by \citet{Palau2021}.

\subsubsection{A total magnetic field estimation}
\label{sse:totB}

Additional insight can be obtained from the results obtained by \citet{Hunter2018}, who
measured the magnetic field on the line of sight,  B$_{\mathrm{los}}$, by detecting the Zeeman effect from OH maser emission yielding magnetic field strengths between +0.5 and +3.7 mG towards the CM2 source and -2 to -5 mG around the MM3 {\HII} region.
Although their resolution, $\sim 520$ au, appears to be larger than the likely physical size of a maser spot, their measurements appears consistent with others \citep{Caswell2011}.
We find our field estimates onto the plane of the sky, B$_{\mathrm{pos}}$, consistent with this range in field strength. The OH maser emission  detected by \citet{Hunter2018} is associated with regions within our map (CM2 and MM3) and with gas at densities in between $10^{5} - 10^{8}$ cm$^{-3}$ 
according to the models of \citet{Cragg2002}, which are within the ranges covered by our density model. This allow us to estimate the total magnetic strength field by computing the vector sum $\mathcal{B}=\sqrt{\mathrm{B}^{2}_{\mathrm{pos}} + \mathrm{B}^{2}_{\mathrm{los}}}$. \citet{Hunter2018} found the distribution of OH masers to be well correlated with methanol masers at 6.7 GHz and with the MM3 UC{\HII} region and CM2 source. However, there seems to be a consistent lack of maser emission towards the strongest continuum peaks in NGC6334I, particularly the MM1 proto-cluster where 80\% of the maser emission is found outside the 40\% level of the continuum in their ALMA 1 mm data. This suggests that the density in the strongest dust emission peaks in MM1 is too high to support inversion \citep{Hunter2018}. Thus, to add the OH Zeeman measurement to our magnetic field map, we decided to take the average of the absolute values from \citet{Hunter2018} Table 8, or $\left< \mathrm{B}_{\mathrm{los}} \right>= 3.4$ mG, and compute the vector sum with the average of our $\mathrm{B}_{\mathrm{pos}}$ values. This yield a total magnetic field estimate $\mathcal{B} = 4$ mG. In principle, one would have liked to add the Zeeman values following a density profile. However, the coverage of the maser emission is sparse and we lack of a well defined canonical shape of the field which might have allow us to produce a total magnetic field map.
This $\mathcal{B}$ estimate is consistent with extrapolations derived by assuming a field strength dependency with density of as obtained from CN Zeeman measurements \citep{Crutcher2019} and by estimations using the DCF method, and its variants, toward other HMSFR \citep{Beltran2019, Cortes2021b, Sanhueza2021}.

\subsubsection{Caveats}
\label{sse:caveats}

The DCF technique has a number of caveats which have been explored in a number of works \citep{Heitsch2001,Crutcher2004,Falceta2008,Cho2016,Skalidis2021,Cortes2019,Liu2021,Lazarian2022,Myers2024}. Here, we mention the most important caveats that might be affecting the validity of our analysis in order to give perspective to the conclusions that we are deriving here. The DCF technique provides an estimate of the field strength onto the plane of the sky by assuming equipartition between turbulence and magnetic energy without considering self-gravitation and the anisotropic nature of turbulence. A region such as NGC6334I has already formed a number of protostars suggesting that the gas is far from an equilibrium situation. 
Although the effect of self-gravity in the dispersion of the position angle is difficult to quantify, gravity manifests itself over long distances requiring large reservoirs of mass to have a significant effect. Thus, the injection
of turbulence by gravity into the gas, which will ultimately  perturb the field, might happen at larger scales than the
one traced by our moving box. Another issue relates to the velocity dispersion obtained from the  C$^{33}$S moment 2 map which   contains contributions from the mean velocity field which in this region might come from the outflow, infalling motions, and (or) rotation, in addition to turbulence. These components cannot be immediately removed because we lack sufficient information about the kinematics and geometry of the source. In short, we might be overestimating the non-thermal motions due to turbulence which will bias the estimation of the field strength as well as the sonic  Mach number (see section \ref{sse:mach} for a discussion).

The estimation of the density model is also subjected to a number of assumptions which are almost impossible to quantify. For instance, the determination of the column density map in our data is not only affected by flux calibration and  temperature model uncertainties, but also by line-of-sight contamination, field selection, and most importantly by interferometric spatial filtering \citep[see ][]{Ossenkopf-Okada2016}. 
Because we constructed the density model assuming a temperature model which may be overestimating the actual dust temperature, the  $\mathrm{B}_{\mathrm{pos}}$ values might be underestimations.
In the case of the dispersion in the field lines, the proxy used here is the dispersion in the linear polarization angle of the ALMA polarized dust emission. Although we are subtracting the estimated error from the data in quadrature and using circular statistics, the true local perturbations to the mean magnetic field lines might be smoothed out by projection effects, such tangling of the field along the line of sight, the different number of turbulent cells in the line of sight, beam smearing \citep[ though the beam obtained sampling our observations traces scales of 0.002 pc close to the dissipation length scale proposed by  ][ of 0.001 pc]{Li2010}, and others, which might produce over-estimations of B$_{\mathrm{pos}}$.

Despite the fact that these caveats are difficult to quantify, we will attempt to produce a error estimate for  $\mathrm{B}_{\mathrm{pos}}$. We do this by simple propagation of errors over equation \ref{eq:Bpos} and by considering error estimates of the physical parameters that we can quantify (see appendix \ref{ap:err_dcf} for a detailed derivation). The error estimates that we obtained are $\sigma_{\mathrm{n}} = 5\times 10^{7}$ cm$^{-3}$ for the number density, $\sigma_{v} = 2$ {\kms} for the velocity dispersion, and $\sigma_{\delta \phi} = 5^{\circ}$ for the position angle dispersion. Using these values, we obtain a mean $\mathrm{B}_{\mathrm{pos}}$ error estimate, $\left< \sigma_{\mathrm{B}} \right> = 1 $ mG after rounding, which yields  $\left< \mathrm{B}_{\mathrm{pos}} \right> = 1.9 \pm 1$ mG. From the values of $\mathrm{B}_{\mathrm{los}}$ obtained by \citet{Hunter2018}, we estimate an error also of 1 mG which after adding them  in quadrature and rounding we obtained  $\left< \mathcal{B} \right>= 4  \pm 1$ mG.

Finally, we should put in perspective that despite these caveats and a simple error estimation, our estimates are consistent with values derived towards other HMSFR which are inline to what we should expect from extrapolations of the Zeeman effect to the densities traced by our data. Finally, in absence of the Zeeman effect, the DCF technique is the simplest method that we have to obtain a estimate of field strength in star forming regions and thus the values derived here  should be considered as such\footnote{The intensity gradient technique is an alternative method which we will consider in subsequent work \citep[see ][]{Koch2012a, Koch2012b, Koch2013}}.

\subsection{Energy Balance}
\label{sse:energy}

To understand the effects of the multiple physical parameters on the magnetic field, we calculate energy  maps associated to turbulence (kinetic energy), gravitational, thermal, and magnetic. 
The kinetic energy map is estimated by computing 

\begin{equation}
K = \frac{1}{2} \int \rho v^{2} d^{3}x
\end{equation}

which for every pixel in the map can be approximated by 

\begin{equation}
K_{i,j} = \frac{1}{2} \mathrm{m}_{i,j} \Delta v_{i,j}^{2}
\end{equation}

where $\mathrm{m}_{i,j}$ is the mass per pixel obtained from the column density map, $\Delta v_{i,j}$ is the velocity dispersion obtained from the C$^{33}$S moment 2 map where we remove the contribution from the thermal sound speed. The dispersion is calculated it as $\Delta v_{i,j} = \sqrt{\Delta v^{2}_{\mathrm{obs},i,j}  - \sigma^{2}_{\mathrm{th}}}$, where $\Delta v_{\mathrm{obs},i,j}$ is the observed velocity dispersion per pixel from the C$^{33}$S moment 2 map,  and $\sigma_{\mathrm{th}} = \sqrt{k_{\mathrm{B}} T_{i,j}}/ \mathrm{m}_{\mu_{\mathrm{H_{2}}}}$ is the sound speed determined from the temperature model (see Figure \ref{fig:enMaps}), where $k_{B}$ is the Boltzmann constant.
The gravitational energy  is calculated as

\begin{equation}
W =\frac{1}{2} \int \rho \Phi_{g} d^{3}x
\end{equation}

where $\Phi_{g}$ is gravitational potential. This expression can also be approximated by

\begin{equation}
W_{i,j} = \frac{1}{2} \mathrm{m}_{i,j} \sum_{k,l} \left( -G \frac{\mathrm{m}_{k,l}}{|r_{i,j}-r_{k,l}|} \right)
\end{equation}

where the indices $(i,j)$ and $(l,k)$ indicate pixel positions in the map, $\mathrm{m}_{i,j}$ is the mass at pixel (i,j), $|r_{i,j}-r_{k,l}|$ is the distance between the mass elements at pixels $(i,j)$ and $(l,k)$ obtained from the column density map. Although the gravitational energy is negative, we plot the absolute value in Figure \ref{fig:enMaps}. Note, the underlying assumption here is that the dust emission is optically thin. If the dust emission is optically thick towards the more massive cores, the estimated gravitational energy will be a lower bound.
The thermal energy can be expressed as 

\begin{equation}
    U = \frac{3}{2} \int n k_{B} T d^{3}x
\end{equation}
where we assume an ideal gas equation of state. As done before, the thermal energy per pixel can be expressed as

\begin{equation}
U_{i,j} = \frac{3}{2} k_{B} n_{i,j} T_{i,j} \delta \mathrm{V}_{i,j}
\end{equation}

where  $n_{i,j}$ is the number density per pixel, $T_{i,j}$ is the model temperature per pixel,
and $\delta \mathrm{V}_{i,j}$ is the pixel volume. We calculate the pixel volume as follows, given 
 $\mathrm{m}_{i,j} = \rho_{i,j}\mathrm{V}_{i,j} = \mu_{\mathrm{H_{2}}} \mathrm{m_{H_{2}}} \mathrm{n}_{i,j} \mathrm{V}_{i,j}$ and using  $\mathrm{m}_{i,j} = \mu_{\mathrm{H_{2}}} \mathrm{m_{H_{2}}} A N_{i,j}$, the pixel volume can be written as $\mathrm{V}_{i,j}=A N_{i,j}/n_{i,j}$, where $A \sim 1000\, \mathrm{au}^{2}$ is the pixel area
\footnote{A square circumscribing each beam has 256 pixels}. This yields the thermal energy per pixel as

\begin{equation}
    U_{i,j} = \frac{3}{2}Ak_{B}N_{i,j}T_{i,j}
\end{equation}

The magnetic field energy density per pixel is calculated as 
\begin{equation}
    M = \frac{1}{8\pi} \int \mathrm{B}^{2} d^{3}x
\end{equation}

following the thermal energy calculation, the magnetic field energy per pixel can be estimated as

\begin{equation}
\label{eq:magEn}
    M_{i,j} = \frac{A}{8\pi} \mathrm{B}_{i,j}^{2} N_{i,j}/n_{i,j}
\end{equation}

where  $\mathrm{B}_{i,j}$ is the magnetic field strength onto the plane of sky estimated using the DCF method, $N_{i,j}$ is the column density per pixel, and $n_{i,j}$ is the number density per pixel (also see Figure \ref{fig:enMaps}).

We also estimate the energy coming from the expanding MM3 cometary H {\small II} region by assuming that the emission is purely free-free which we use to estimate its thermal energy. \citet{Sadaghiani2020} obtained continuum emission with ALMA at 87.6 GHz. From these data, we estimate the size of the MM3 cometary H {\small II} by a Gaussian fit to their data which gives us a region of $3^{\prime\prime}.3 \times 2^{\prime\prime}.5$ in size. The electron temperature, T$_{e} = 10^{4}$ K,  and density, n$_{e} = 3 \times 10^{6}$ cm$^{-3}$ are taken from  \citet{Brogan2016A} simple free-free model. Thus using these values and assuming the geometrical thickness used for our density model, we obtain $\mathcal{E}_{\mathrm{H\ {\tiny II}}} \sim 10^{45}$ [ergs].

The energy balance  against the magnetic field is calculated by subtracting the turbulent, thermal (both gas and UC {\HII}), and gravitational energies to the magnetic field energy as

\begin{equation}
\Delta E = M - (U + W + K+ \mathcal{E}_{\mathrm{H\ {\tiny II}}} + \mathcal{E}_{\mathrm{outflow}})
\end{equation}

where $\mathcal{E}_{\mathrm{outflow}}$ corresponds to the combined energy in the red and blue lobes of the outflows. The results of the energy balance are listed in Table \ref{table:energybalance}. We do not compute values for MM7 and MM9 because we do not have sufficient  C$^{33}$S coverage over these cores to obtain the velocity dispersion. 
From these results, it is clear that there is sufficient energy in the system to perturb the magnetic field in NGC6334I, even when considering the large uncertainties introduced by our assumptions. 
The differences in energy are about two orders of magnitude between the field and the combined effects of the other forces  around the main cores and in the whole of NGC6334I. The gravitational energy is clearly the dominant factor in the energetics of the region, which is expected given its evolutionary stage. 
Nonetheless, the energy in the outflows alone are an order of magnitude larger than the magnetic energy and because  the outflows might be injecting turbulence at smaller scales than gravity, 
the outflow feedback might be a significant factor when considering the perturbation to the magnetic field morphology.  
Interestingly, the energy in the cometary UC {\HII} region, a single UC {\HII} region, is comparable to the bulk of the thermal energy over the whole of NGC6334I. Thus, the expansion of this UC {\HII} region might also inject additional turbulence in the gas at similar scales than the outflows. Thus, protostellar feedback maybe the dominant driver behind the injection of turbulence in NGC6334I at the scales sampled by our data.
Finally, the analysis done here suggests that a magnetic field in order of milli-Gauss appears not to be a dominant factor at the core scales in NGC6334I.

\begin{figure*} 
\includegraphics[width=0.8\hsize]{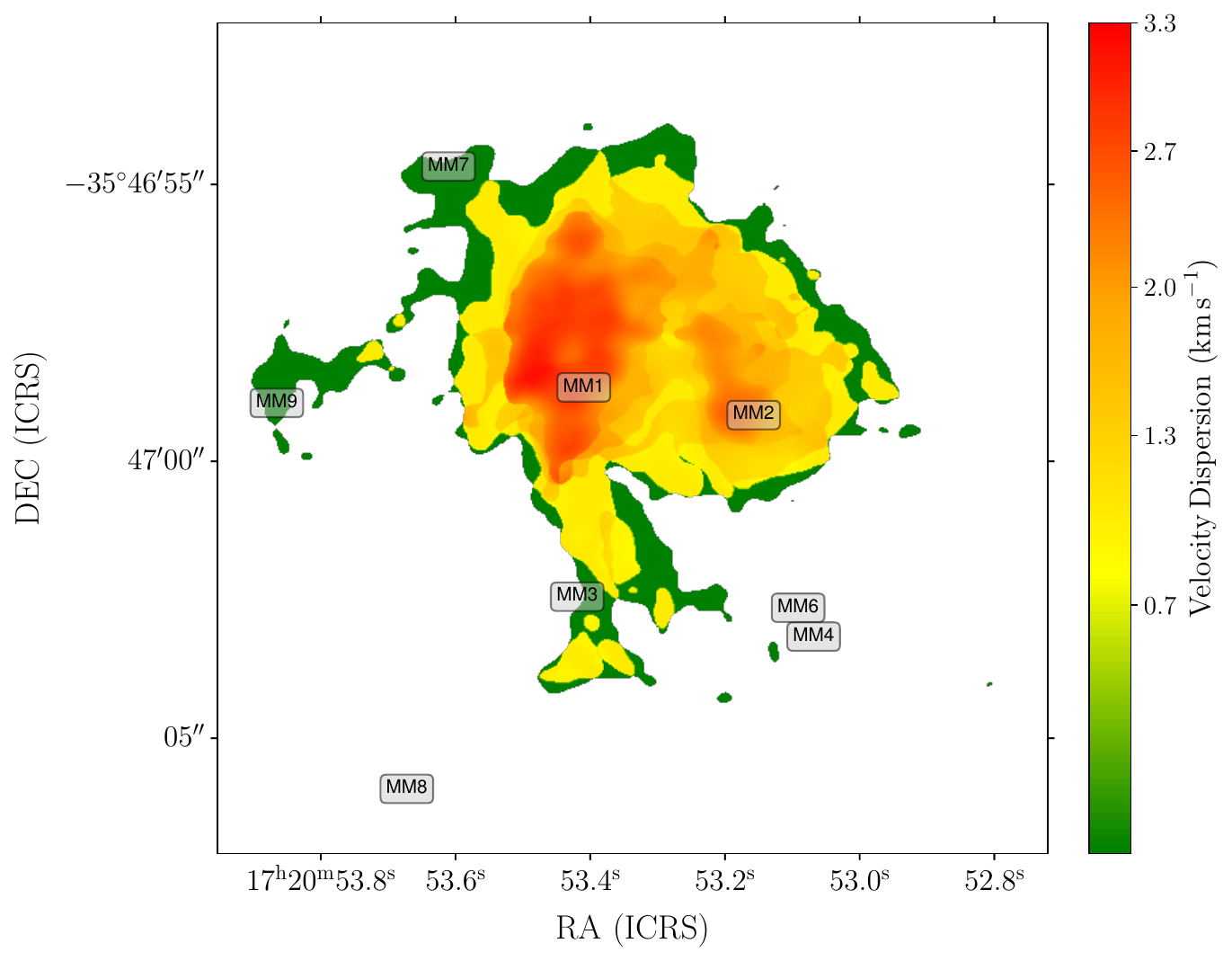}\\
\includegraphics[width=0.8\hsize]{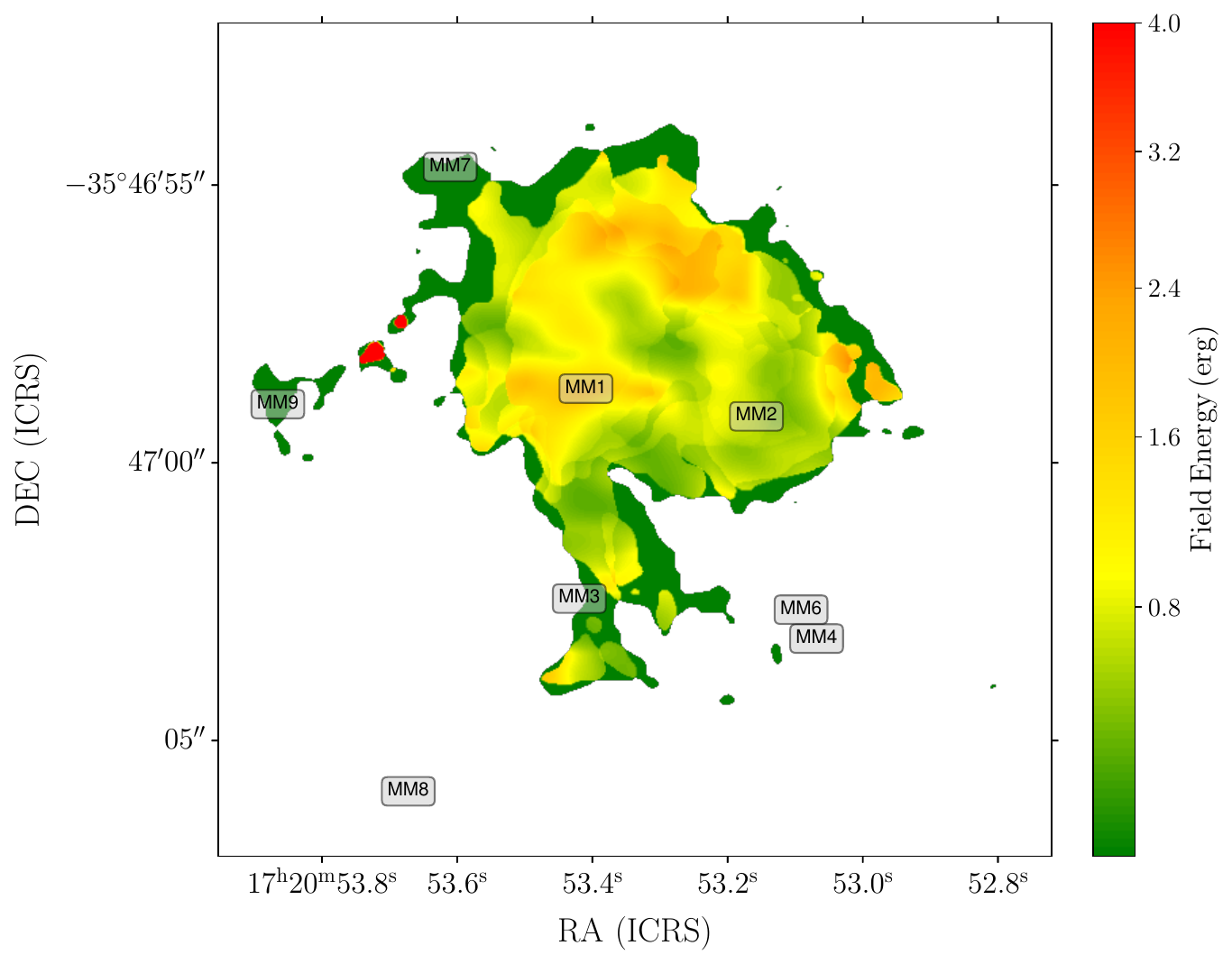}
\smallskip
\caption{ {\bf\em Upper panel}. The moment 2 map from the C$^{33}$S emission from NGC6634I is presented here. The color scale indicates the velocity dispersion in {\kms} where no correction from the mean velocity field has beem applied.  {\bf\em Lower panel}. The Alfven speed map, as a color scale, calculated from the magnetic field strenght onto the plane of the sky and our number density model is presented here. 
\label{fig:vdisp}
}
\bigskip
\end{figure*}

\subsection{Effect of the Field in the Dynamics of the Gas}
\label{sse:mach}

To study the effect of the magnetic field in the gas dynamics, one can start by comparing the different speeds relevant to 
the physical processes affecting the gas.
 These are the sound speed, the non-thermal velocity dispersion, obtained from optically thin molecular emission, and the Alfven speed, which we compute from estimates of the magnetic field strength. Both the velocity dispersion and the Alfven speed can be used as a proxies for the different modes of hydromagnetic waves which encompass what we understand as MHD turbulence \citep{Mouschovias2011}. Although complex in nature, in principle we can compare thermal to kinetic by computing the Mach number, or $\mathcal{M}_{s}$, and kinetic to magnetic by computing the Alfvenic-Mach number, or $\mathcal{M}_\mathrm{A}$. For supersonic motions, non-thermal motions dominates the dynamics where thermal energy only becomes relevant at larger densities. However if the magnetic field is strong enough, we can have a sub-Alfenic regime, where $\mathcal{M}_\mathrm{A} < 1$ and $\rho V^{2} \leq B^{2}$, which is still supersonic.
In this case the magnetic field can significantly influence the dynamics through the Lorentz force \citep{Beattie2020}. In the super-Alfvenic regime where $\mathcal{M}_\mathrm{A} > 1$, the magnetic field plays a lesser role, but different hydromagnetic wave modes still exist which can influence the gas dynamics \citep{Mouschovias2011}.

Preliminary studies done with the SMA, Herschel, and APEX over the whole of the NGC6334 molecular cloud complex, suggested that ($\mathcal{M}_{s} \sim 6$), but sub-Alfvenic ($\mathcal{M}_{A} \sim 0.9$) at resolutions of $\sim 0.2$ pc \citep{Zernickel2015}. Furthermore, \citet{Li2015} estimated the magnetic field onto the plane of the sky to conclude that the gas motions in NGC6334 are also likely to be sub-Alfvenic. Recently, 
ALMA observations of the NGC6334S IRDC (also part of the NGC6334 complex) used
 emission from H$^{13}$CO$^{+}$ and NH$_{2}$D, to explore the non-thermal motions in the IRDC at 0.02 pc resolution with ALMA \citep{Li2020}. They found that the spatially unresolved non-thermal motions are predominantly subsonic and transonic ($\sim$ 77\% regions), but they made no estimations about the Alfvenic regime of the filament. The sub to transonic dominated regime in  the NGC6334S IRDC suggests that the region is  at a very early evolutionary stage such that the protostellar feedback-induced turbulence is small as compared to the initial turbulence \citep{Li2020}.
 This findings were confirmed by \citet{Liu2023a} who explored the evolution of the sonic Mach number from large scales ($\sim 15$ pc resolution) to small scales ($\sim 0.005$ pc resolution) finding the gas is mostly supersonic over 3-4 orders of magnitude in length-scales over most of NGC6334, but sub-sonic to trans-sonic in NGC6334S (the IRDC), which is consistent with what has already been observed in other IRCDs \citep{Sanhueza2017}. This IRDC is at an earlier stage of evolution respect to NGC6334I, but because is part of the same molecular complex, the cores in NGC6334S may indicate what were the initial conditions in NGC6334I.

We investigate the non-thermal motions in  NGC6334I to explore how much influence the magnetic field has in the gas dynamics. To achieve this, we compute the sound speed map using the temperature model map, represented as $c_{\mathrm{s}} = \sqrt{k_{\mathrm{B}}T/\mathrm{m}_{\mathrm{H}_{2}}}$. Employing the velocity dispersion map derived from C$^{33}$S
(see section \ref{sse:energy}), we estimate the sonic Mach number map using $\mathcal{M}_{s}=\sqrt{3}c_{\mathrm{s}}/\Delta V$ (see Figure \ref{fig:mach}).
The Mach number map suggests that the gas is supersonic throughout most of the regions covered by the C$^{33}$S emission in NGC6334I. Mach number values as high as $\mathcal{M}_{s} \sim 4$ suggests that non-thermal motions are high with turbulence 
being injected at the core scales to sustain the gas velocity dispersion seen here.
Our values for the sonic Mach number are consistent with the previous results and with a number of other evolved high-mass star forming regions  \citep{Pattle2023,Pineda2023}.
To estimate the Alfvenic regime, we calculate the
Alfven speed  $V_{\mathrm{A}} = \mathcal{B}/\sqrt{4\pi\rho}$ but using $\mathrm{B_{pos}}$ as an estimate of $\mathcal{B}$. The $V_{\mathrm{A}}$ corresponds to the propagation speed of the transverse waves  along the magnetic field lines. These motions do not involve compressions or rarefactions of the gas, meaning there is no density variation associated with these waves (different from the fast and slow magnetosonic modes). 
Interestingly, when using solely $\mathrm{B_{pos}}$ as the $\mathcal{B}$ estimate, the Alfven speed does not depend on the density. This is because
$V_{\mathrm{A}} = \mathrm{B}_{\mathrm{pos}}/\sqrt{4 \pi \rho} = \delta V / \delta \phi $ when introducing the original DCF equation, $\mathrm{B}_{\mathrm{pos}} =\sqrt{4 \pi \rho} \delta v/ \delta \phi$, into the Alfven speed expression. 
Because DCF assumed equipartition over the total magnetic field, a derivation of the Alfven speed based on the DCF alone is as good as the $\mathrm{B_{pos}}$ estimate and subjected to the same caveats. We show the Alfven speed map in Figure \ref{fig:vdisp}. Estimates of the Alfven speed from the literature are scarce. However, \citet{Roshi2007} estimated Alfven speeds ranging between 0.7 and 4 {\kms} based on Carbon recombination line observations conducted with the Arecibo telescope. These observations targeted a set of 14 PDRs surrounding {\HII} regions associated with high-mass star-forming regions. \citet{Liu2020} estimated Alfven speeds from their polarization mapping of the G28.34+0.06 IRDC, finding values between 0.5 to 1.5 {\kms}. The values from both studies  are consistent with our findings (see Figure \ref{fig:vdisp}).

From the Alfven speed map alone it is difficult to explore the effect of the field in the gas. To this effect, we can compute the Alfven Mach number map using $\mathcal{M}_{\mathrm{A}} = \Delta V/V_{\mathrm{A}}$ (also see Figure \ref{fig:mach}). As with the Alfven speed, we note that the Alfven Mach number depends only on the dispersion of the magnetic field lines when assuming DCF, which gives $\mathcal{M}_{\mathrm{A}} \sim \delta \phi$.
We found this approach convenient because, in principle, it allows for an estimate of this quantitative from the polarization map alone. Furthermore, the Alfven Mach number map derived here is not affected by the assumptions used to derive the temperature and density models. Furthermore, even by considering different variants of DCF, the Alfven Mach number will still strongly depend on the dispersion of the magnetic field lines.  
The Alfven Mach number map shown in Figure \ref{fig:mach}, strongly suggests that most of NGC6334I appears to be in trans-Alfvenic conditions
($\mathcal{M}_{\mathrm{A}} \sim 1$) evolving to super-Alfvenic conditions ($\mathcal{M}_{\mathrm{A}} > 1$) as we move into the cores. 
Although there are regions at the edges of the map which may be sub-Alfvenic ($\mathcal{M}_{\mathrm{A}} < 1$), their extension 
is small and thus inconclusive given the area covered by the C$^{33}$S mask. 
In trans-Alfvenic conditions, the non-thermal motions can be considered to be in equipartition with the magnetic field, suggesting that the effect of the magnetic field and MHD turbulence are comparable. As density increases due to the gravitational
pull, the Alfven Mach number becomes super-Alfvenic where the magnetic field is dominated by non-thermal motions. This becomes evident over the peaks with $\mathcal{M}_{A} \sim 4$, where the exception is MM4 where the C$^{33}$S emission is marginal (see Figure \ref{fig:c33s_moms}).
These results agree well with the study by \citet{Pattle2023} who found that the turbulence is mostly trans-Alfvenic when statistics are compiled over many sources.
The strongest peaks in the Alfven Mach number map also correlate well with the regions where we see $\sim 90^{\circ}$ deviations in the field orientation within scales of 1$^{\prime \prime}$. 
Note, if the region were sudden $\sim 90^{\circ}$ turns are the product of projection effects, then the estimated dispersion in the field should decrease the dispersion in the field lines increasing the estimated value for B$_{\mathrm{pos}}$ at those regions. Conversely,  this should also decrease the value of $\mathcal{M}_{A}$ in those regions as well.
The transition seen from trans-Alfvenic to super-Alfvenic in NGC6334I is consistent with an evolved HMSFR such NGC6334I where gravity dominates the gas dynamics and the magnetic field appears to be dynamically important only at the edges of the region.
Interestingly, new high angular resolution ($\sim 0^{\prime\prime}.065$) ALMA results from the ``hourglass'' magnetic field in  G31.41+0.31 suggests that the gas can  evolve from sub-Alfvenic to super-Alfvenic in a progression resembling NGC6334I, if we were to assume that the edges of our Alfvenic Mach number map correspond to a sub-Alfvenic regime \citep{Beltran2024}. 
Now, when considering the whole NGC6334 molecular complex, exploring the Alfvenic state on NGC6334I(N), a source supposedly in an earlier stage of evolution, might show a more conclusive sub-Alfvenic to super/trans-Alfvenic condition from the outer to inner regions of the clump while the NGC6334S IRDC might likely be sub-Alfvenic. This would certainly give us significant insight about the evolutionary effect of the magnetic field over the gas dynamics in NGC6634.  We leave this for future work.

\begin{figure*} 
\includegraphics[width=0.49\hsize]{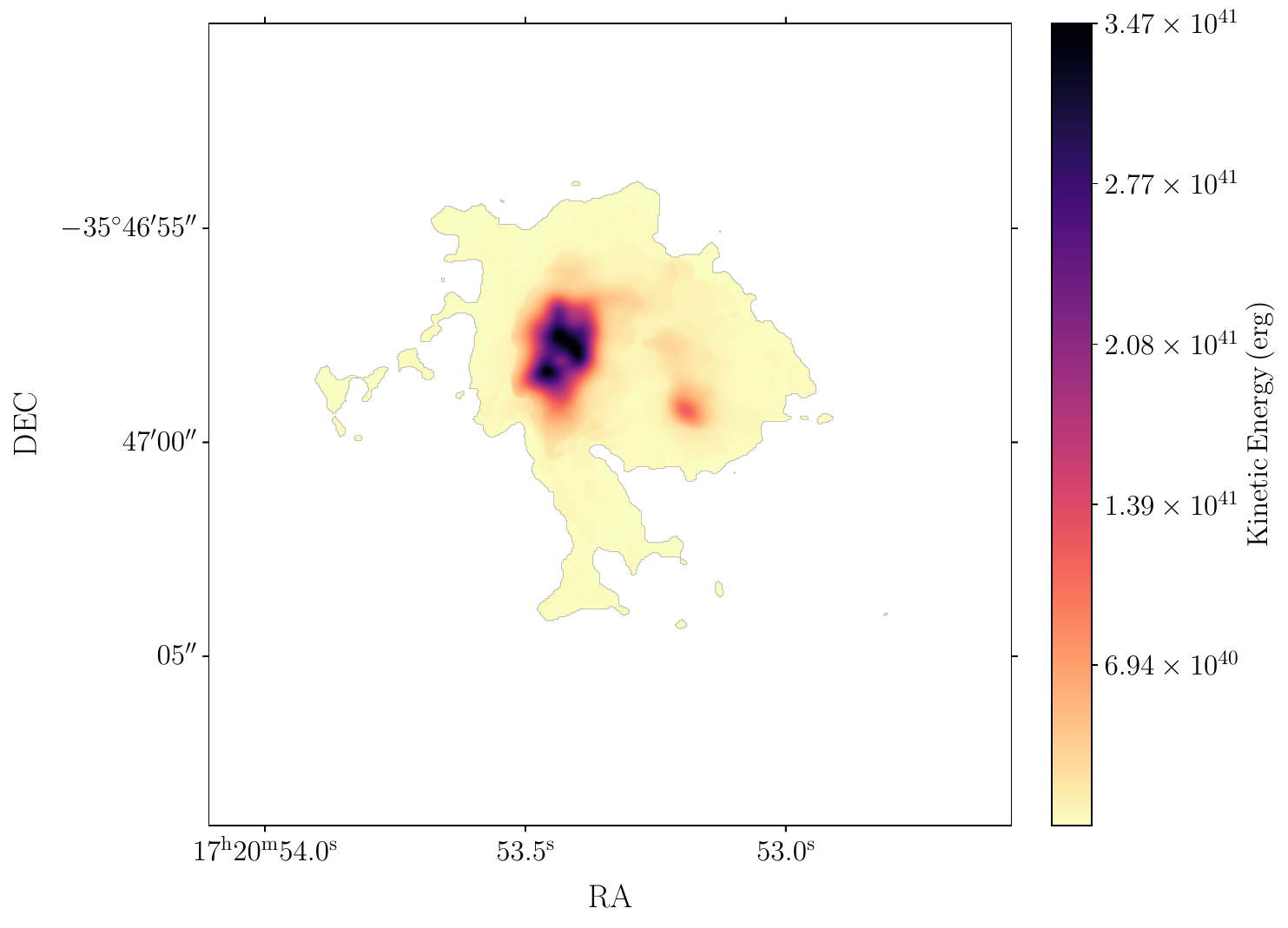}
\includegraphics[width=0.49\hsize]{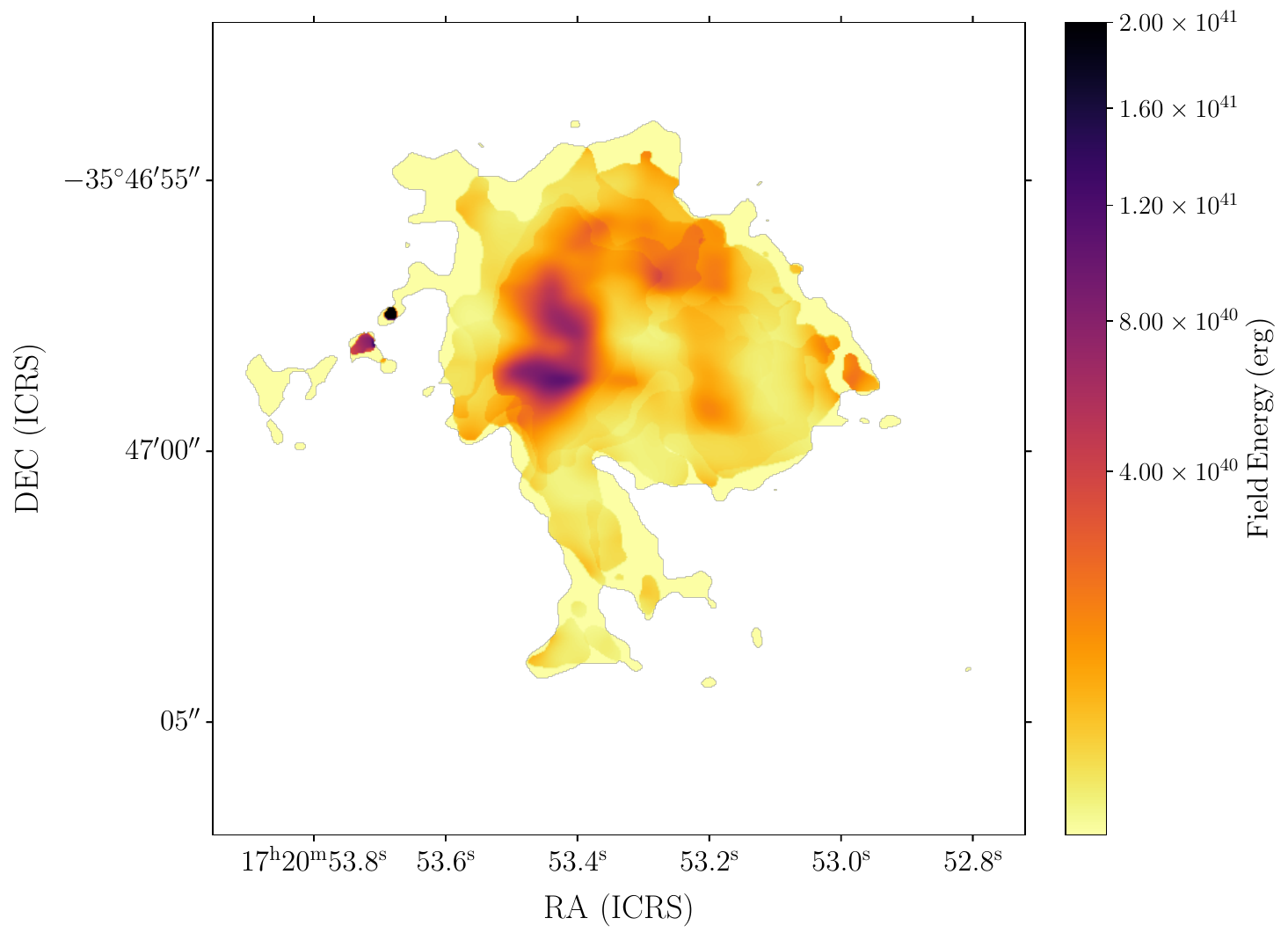} \\
\includegraphics[width=0.49\hsize]{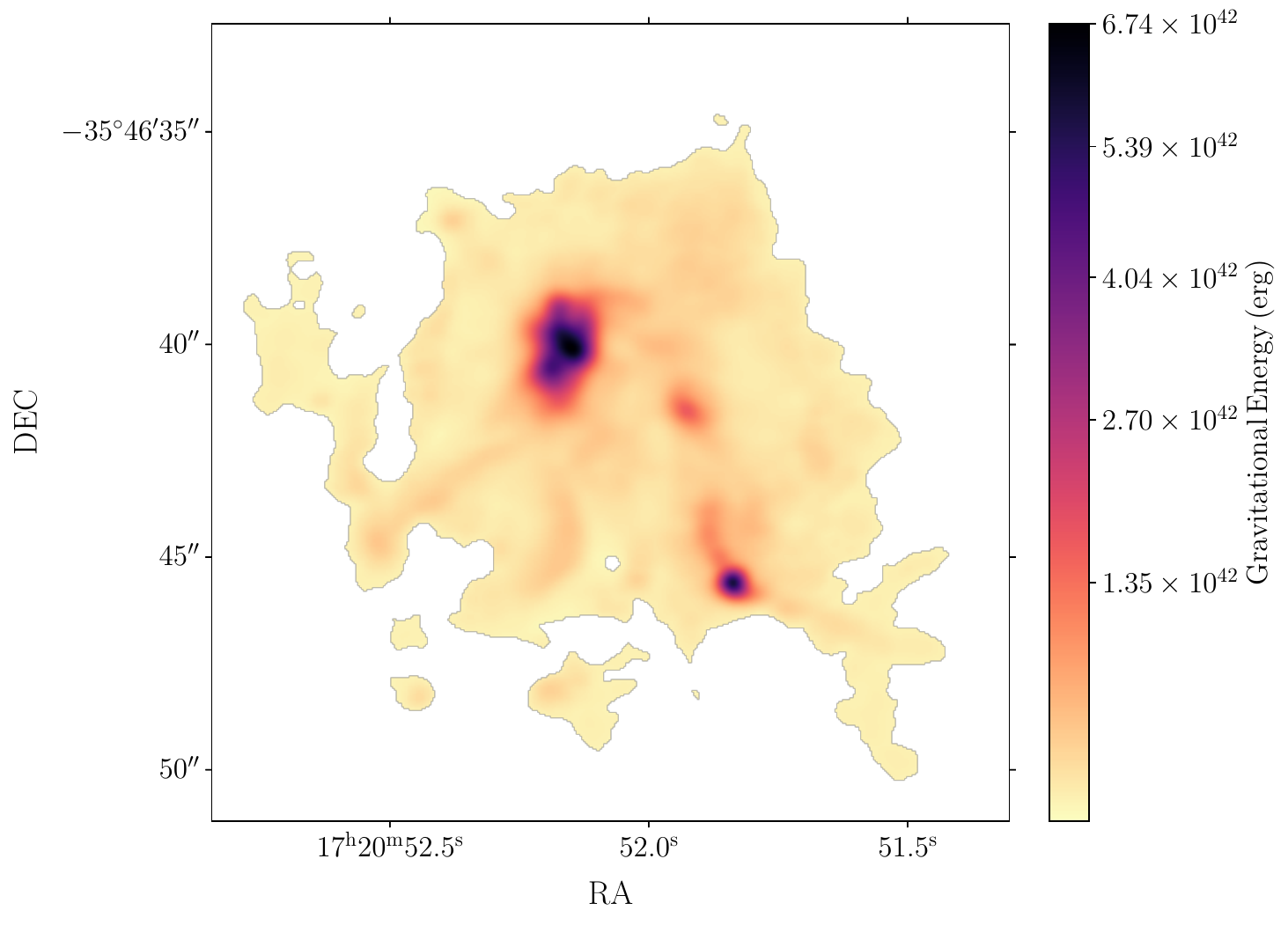}
\includegraphics[width=0.49\hsize]{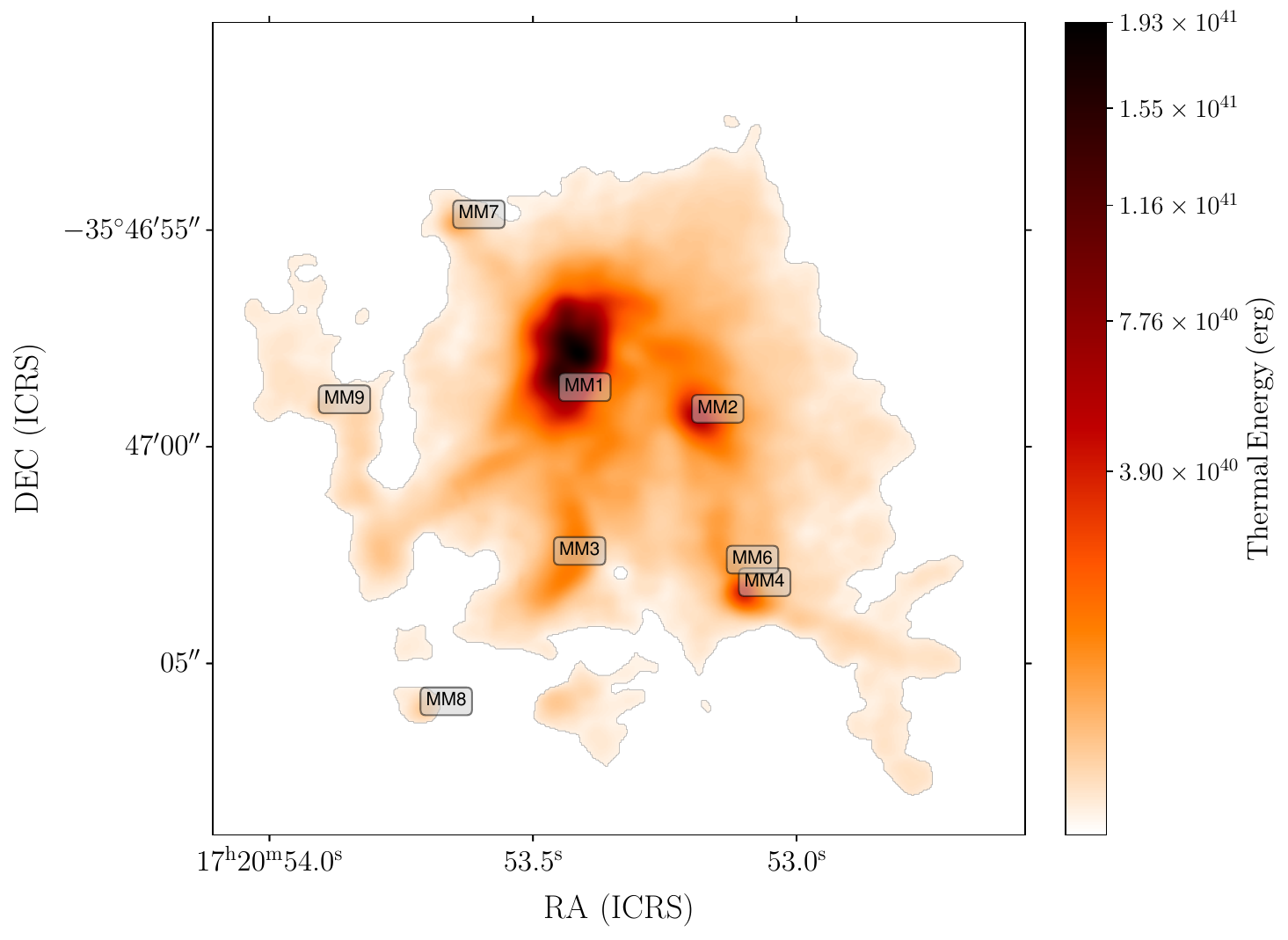}
\smallskip
\caption{{\bf\em Upper panel.} The kinetic and magnetic  energy maps are shown here. The extent of the maps is given by the C$^{33}$S moment 2 map which is smaller than extend of the dust continuum (see Figure \ref{fig:c33s_moms}). 
{\bf\em Lower panel.} The magnetic field and thermal  energy maps are shown here. Their extension is given by the dust continuum map  which wass used to define the temperature model (see section \ref{se:dust}). The source labels are shown over the thermal energy map only to improve visual inspection.
\label{fig:enMaps}
}
\bigskip
\end{figure*}

\begin{figure*} 
\centering
\includegraphics[width=0.8\hsize]{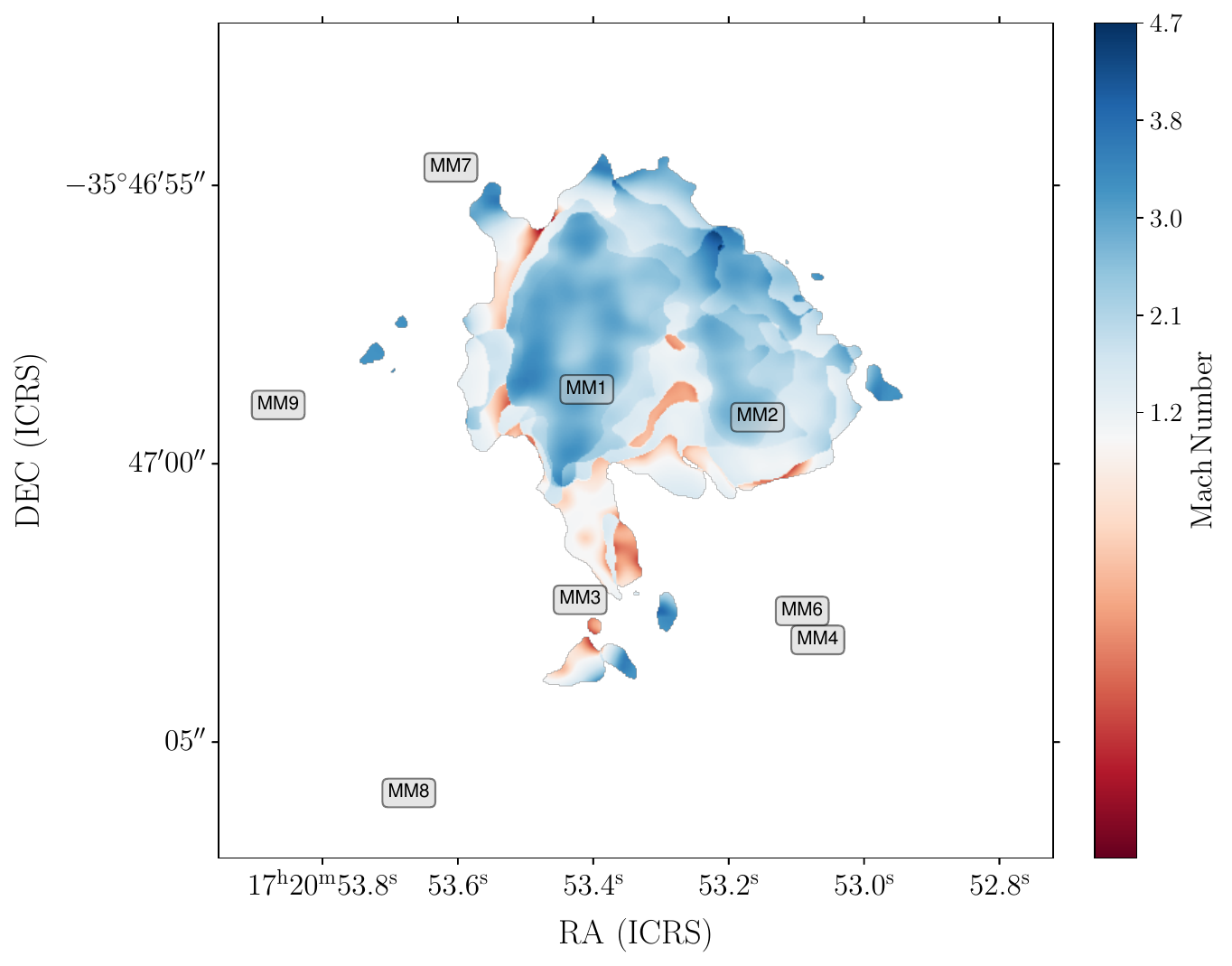}\\
\includegraphics[width=0.8\hsize]{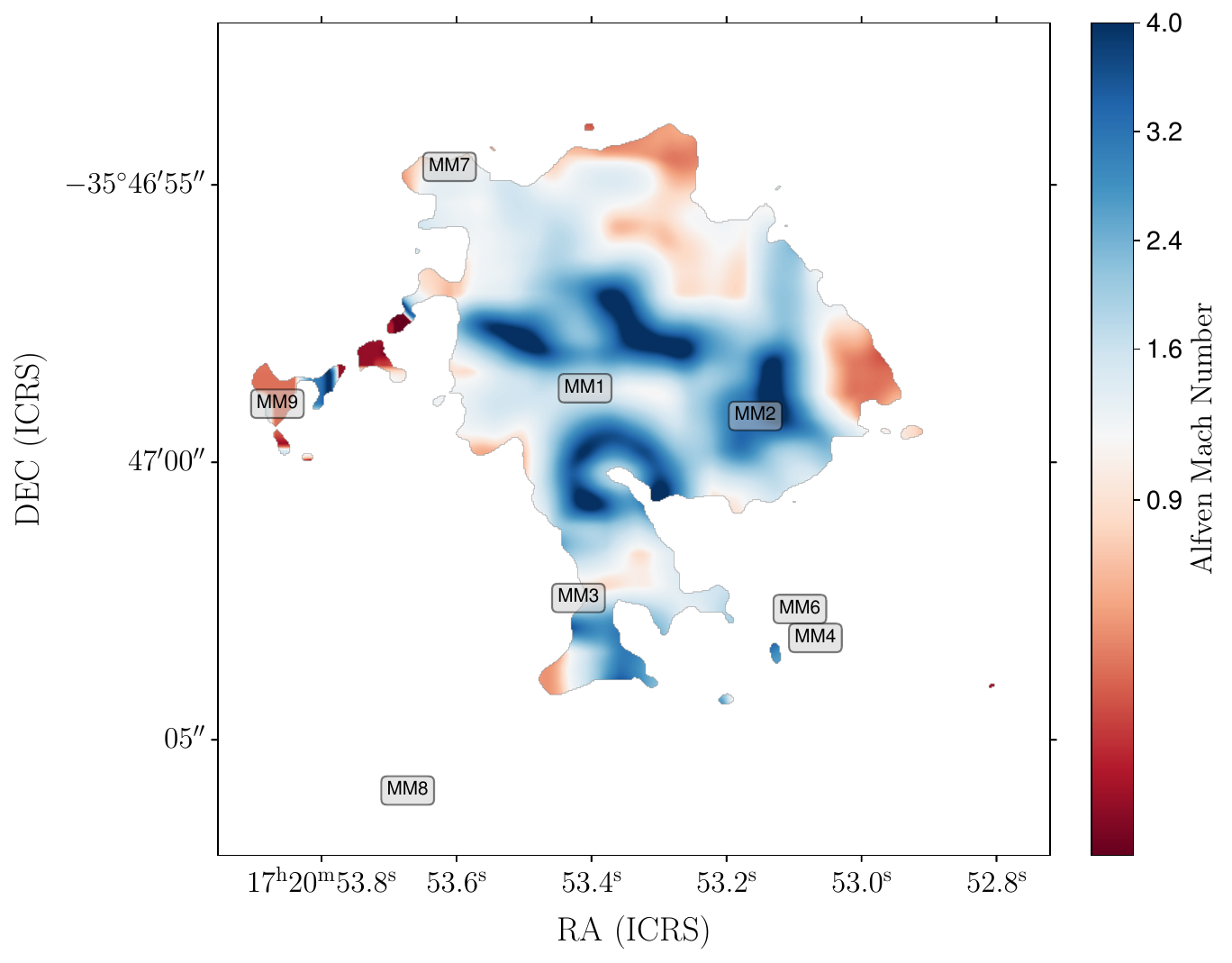}
\smallskip
\caption{ {\bf\em Upper panel.} The sonic Mach number map, $\mathcal{M}_{s}$, is shown in color scale.   {\bf\em Lower panel.} The Alfven-Mach number map, $\mathcal{M}_\mathrm{A}$, is also shown in color scale where trans-Alfvenic conditions are represented by color values $\sim 1$.
\label{fig:mach}
}
\bigskip
\end{figure*}

\section{SUMMARY AND CONCLUSIONS}
\label{se:conc}

We have presented sub-arcsecond resolution results from polarized dust emission and total intensity line emission observations, at 250 GHz, towards NGC6334I with ALMA. The results can be summarized as follows:

\begin{itemize}

\item The morphology of the magnetic field onto the plane of the sky, was derived from 1.2 mm polarized dust emission under the assumption of grain alignment by magnetic fields. The magnetic field morphology is in close agreement with the ALMA results from \citet{Liu2023} taken at 1.4 mm. 

\item Our two-epoch data over five months show no substantial change in total intensity ($< 1\%$) or linear polarization ($< 8\%$), indicating a stable period or the end of the outburst.

\item To explore the effect of the four outflows detected from NGC6334I in the magnetic field, we quantified their energy by making used of the CS emission. We found that the bulk energy in the outflows is $\sim 3.5 \times 10^{45}$ ergs. 

\item By making used of the temperature model produced by \citep{Liu2023}, we computed energy maps for the kinetic, thermal, gravitational, and magnetic field, whose strength onto the plane of the was estimated using the DCF technique obtaining values between 1 to 11 mG with a mean magnetic field strength of 1.9 mG. When we add Zeeman measurements from OH maser observations \citep{Hunter2018}, we estimate a mean total magnetic field strength of 4 $\pm 1$ mG in agreements with others.

\item While the magnetic field in NGC6334I holds considerable energy, our analysis reveals that the cumulative effect of kinetic, thermal, and gravitational energies, alongside the contributions from outflows and the external pressure exerted by the surrounding cometary {\HII} region, is capable of overpowering the magnetic field. 
Furthermore, the energy from the outflows alone may sufficiently disturb the field lines through the injection of turbulence at the core scales. At the same time, gravity-induced turbulence is likely to occur on larger scales. Therefore, protostellar feedback potentially emerges as the primary influence on the observed perturbations in the magnetic field's structure across NGC6334I.

\item By making use the optically thin C$^{33}$S emission, we computed a velocity dispersion map which we use as a proxy for the non-thermal motions. Also, by using the temperature map, we computed the thermal sound speed map and the sonic Mach number map. We found that the gas is mostly supersonic throughout NGC6334I consistent with previous findings. In the same way, we estimated the Alfven speed which we used to compute the Alfven Mach number. From the Alfven Mach number map, we found indications suggesting that gas smoothly evolve from trans-alfvenic to super-Alfvenic, with possibly sub-Alfvenic regions when we consider the outer edges of the map. This suggests that we may be seeing the progression at which the field finally gets overwhelmed by the effects of proto-stellar feedback induced turbulence and gravity in high-mass star formation.

\end{itemize}

\setlength{\tabcolsep}{2pt}
\begin{deluxetable*}{cccccccc}
\tablecolumns{8}
\tablewidth{0pt}
\tabletypesize{\scriptsize}
\tablecaption{ Energy Balance \label{table:energybalance}}
\tablehead{
\colhead{Region} &
\colhead{Kinetic} &
\colhead{Thermal Gas} &
\colhead{Thermal {\tiny H II}} &
\colhead{Gravitational} &
\colhead{Outflow} &
\colhead{Magnetic} &
\colhead{Balance}  \\
\colhead{}     &
\colhead{$10^{46}$ (erg)}     &
\colhead{$10^{46}$ (erg)}     &
\colhead{$10^{46}$ (erg)}     &
\colhead{$10^{46}$ (erg)}     &
\colhead{$10^{46}$ (erg)}     &
\colhead{$10^{46}$ (erg)}     &
\colhead{$10^{46}$ (erg)}      
}
\startdata
MM1 & \phantom{0}0.147 & \phantom{0}0.025 & - &\phantom{0}0.762 & - & \phantom{0}0.016 &-0.918 \\
MM2 & \phantom{0}0.214 & \phantom{0}0.040 & - &1.3 & -& 0.037 & -1.485 \\
MM4 & \phantom{00}0.013 & \phantom{00}0.002 & - &\phantom{0}0.070 & - & \phantom{0}0.002 & \phantom{0}-0.082 \\
NGC6334I & \phantom{0}0.538 & \phantom{0}0.100 & 0.14 & 3.163 & 0.350 & 0.087  & -3.714 
\enddata
\vspace{0.1cm}
\tablecomments{The energies are calculated by summing all the pixels inside the region and balance is calculated by subtracting all the energies to the magnetic field energy. The region labeled NGC6334I corresponds to the whole total intensity map. The energy in the expanding cometary UC {\tiny H II} region and the outflows are considered only for the bulk of NGC6334I.}
\end{deluxetable*}

\appendix
\section{Comparison with lower frequency polarization results}
\label{ap:fcomp}

\begin{figure*} 
\includegraphics[width=0.95\hsize]{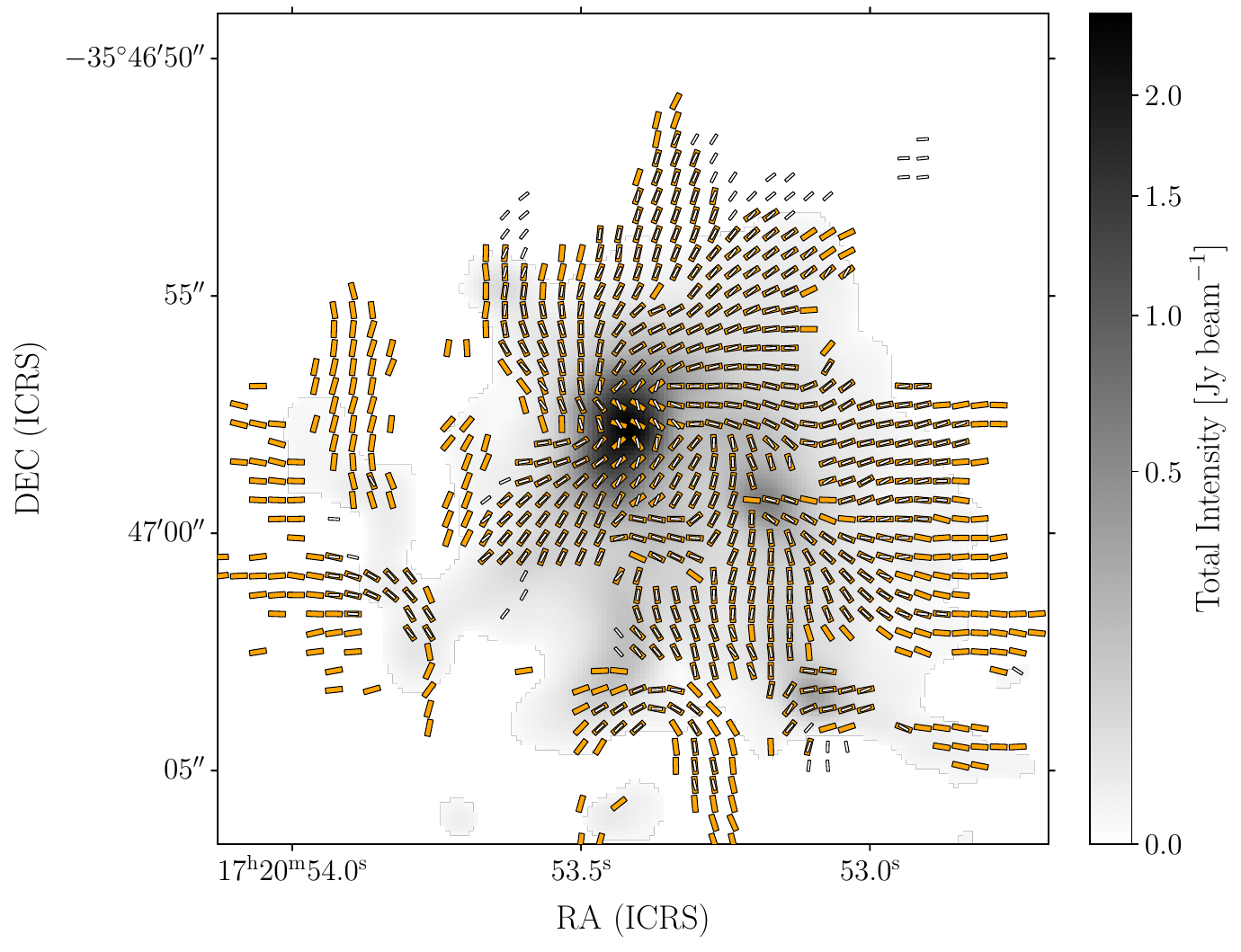}
\smallskip
\caption{Magnetic field morphology in NGC6334I derived from polarized dust emission at 220 GHz (white pseudo-vectors) and 250 GHz (orange
pseudo-vectors). The gray scale corresponds to the Stokes $I$ from 250 GHz dust emission. The white pseudo-vectors were plotted using a smaller length and width to allow comparison between the two different field morphologies.
\label{fig:NGC6334I_COMP}
}
\bigskip
\end{figure*}

\begin{figure*} 
\includegraphics[width=0.95\hsize]{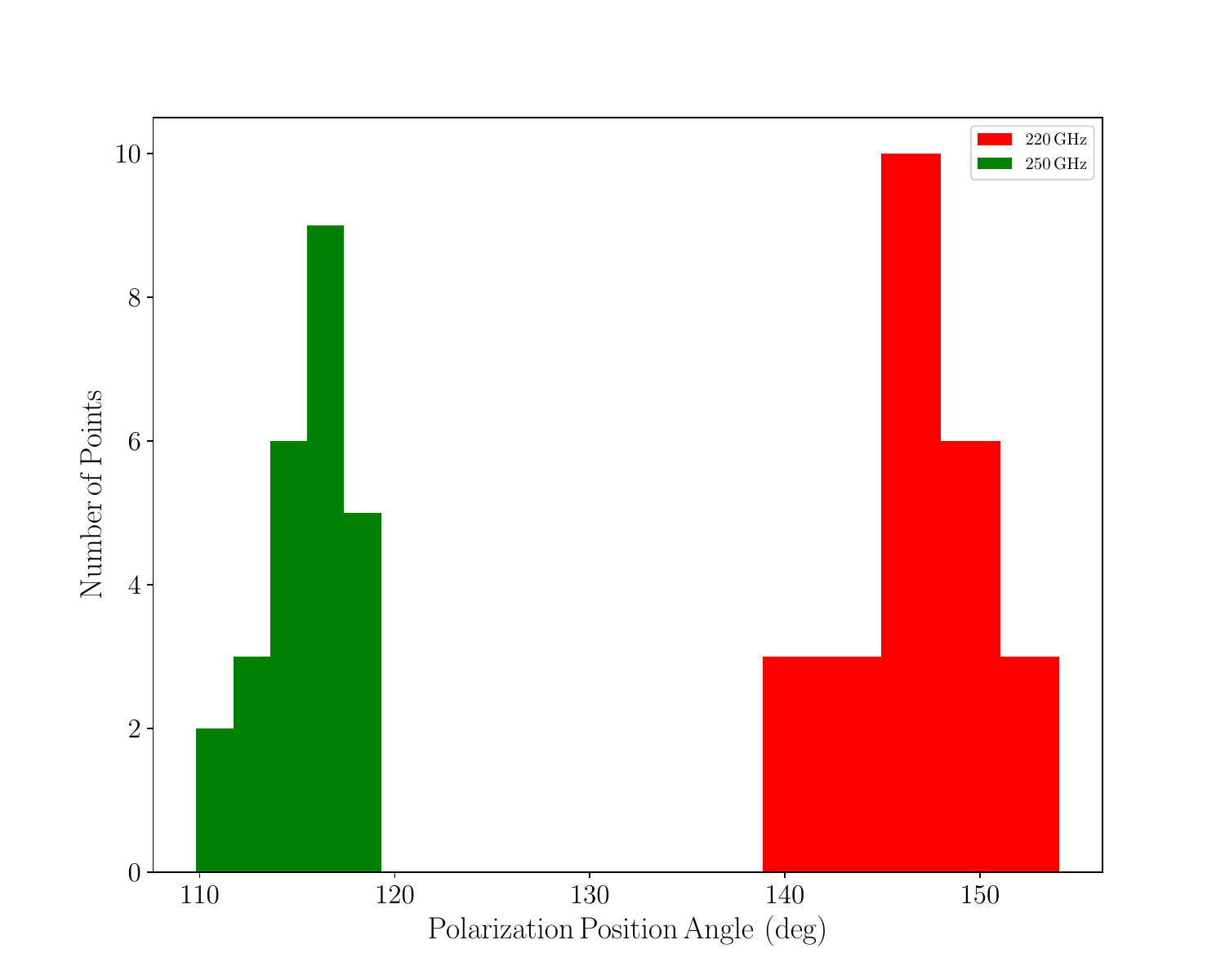}
\smallskip
\caption{The histogram shows the position angle values between the 220 GHz \citep{Liu2023} and the 250 GHz (ours). A clear separation of $\sim 35^{\circ}$ is seen from the histogram.
\label{fig:polaDelta}
}
\bigskip
\end{figure*}

\begin{figure*} 
\includegraphics[width=0.95\hsize]{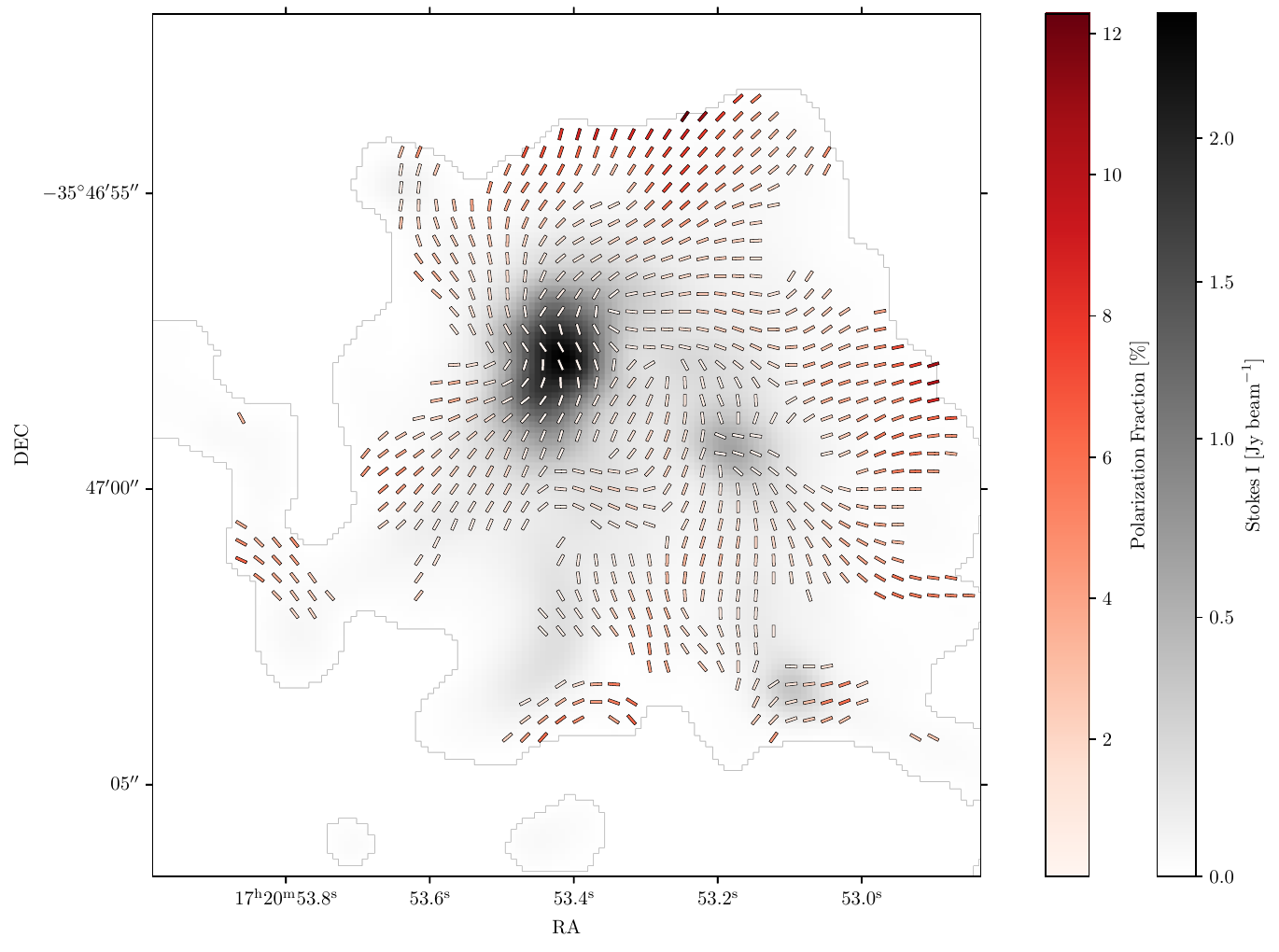}\\
\includegraphics[width=0.95\hsize]{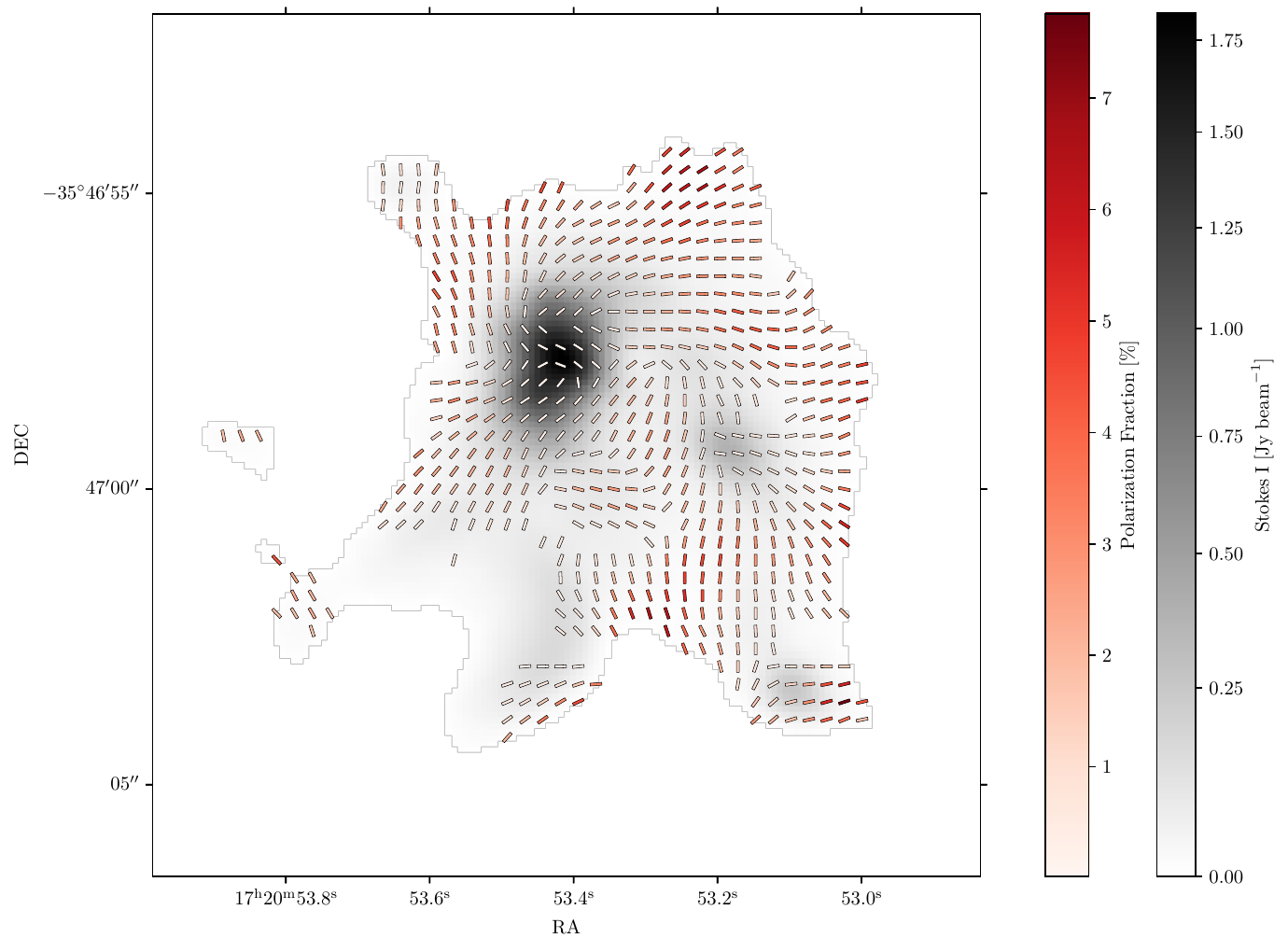}\\
\smallskip
\caption{The magnetic magnetic field morphology maps are show here for both 220 GHz (upper panel) and 250 GHz (lower panel). The color-scale was use in the pseudo-vectors and indicates fractional polarization values while the gray-scale indicates Stokes I (total intensity) emission. 
\label{fig:pfracComp}
}
\bigskip
\end{figure*}

As previously mentioned, the magnetic field morphology in NGC6334I at comparable resolutions but lower frequencies, has already been explored 
with ALMA \citep{Liu2023}. Here we compare our results to those of \citet{Liu2023}, where we see significant consistency 
between the field morphology obtained at 250 GHz (our data) respect to the results from 220 GHz \citep{Liu2023}. Figure \ref{fig:NGC6334I_COMP} shows the superposed field morphologies obtained independently by both datasets. Our dataset was re-gridded to the frame and
resolution of the 220 GHz data ($0^{\prime \prime}.67 \times 0^{\prime \prime}.52$) to allow a direct comparison. 
When inspecting Figure \ref{fig:NGC6334I_COMP}, it is clear that the field morphologies are remarkably similar over most of the extent of the Stokes I emission. The most significant deviation between both datasets are seen at the center of the MM1 proto-cluster. The 250 GHz data shows a North-South orientation while the 220 GHz data presents a more East-West orientation. This difference in the field morphology is significant ($\Delta \phi \sim 35^{\circ}$, see Figure \ref{fig:polaDelta}) when compared to the rest of the map where the pseudo-vectors agree quite well. To explore if this is the result of calibration uncertainties, Figure \ref{fig:pfracComp} shows the derived magnetic field morphology at both frequencies, but now the
pseudo-vectors are colored to indicate the level of fractional polarization (the fractional polarization maps have $3\sigma$ cutoff in Stokes I and a coarser sampling rate). At center, where the Stokes I peaks and the difference in angle is the largest, the range of fractional polarization is between 0.1\% and 0.15\% in both datasets.
Although this range is close to the polarization calibrationa accuracy of ALMA \citep[0.1 \% ][]{Cortes2023}, it is sufficently large to suggests that the difference might be real.
Given the current datasets, it is difficult  to ascertain what is the cause of such difference, particularly considering the excellent overall agreement between them throughout the region where both datasets are separated by just 30 GHz.
Polarized emission by self-scattering appears implausible as an explanation (as discussed in section \ref{sse:fmorph}). High resolution ALMA observations in full Stokes should be conducted to explore this discrepancy further.

\section{The Magnetic Field  Dispersion Map}
\label{ap:std}

To estimate the magnetic field strength onto the plane of the sky using DCF, estimating the dispersion in the polarization position angle is critical. Because we assume flux freezing, the local dispersion in the field lines is produced by collisions between the charge carriers and the neutrals. These collisions perturb the mean field yielding deviations in its local direction, which we understand as the dispersion. The dispersion is an statistical quantity and thus we will use statistical arguments to estimate its value. Because we are interested in obtaining a map, we make use of a moving window to calculate the standard deviation using semicircular statistics.
To define the size of the window, we assume that the angles follow a Gaussian distribution. Although Gaussian distribution is a ``usual'' assumption in astronomy, its usage here can be justified from the point of view of the central limit theorem. If we assume that the perturbations over the field lines are local then each independent point in the map can be considered as a random variable following it own distribution due to the actions of the local turbulence. In the limit of large numbers distribution of the mean of all of independent points follows a Gaussian distribution. 
Moreover, local turbulence can also be modelled using a Gaussian distribution as done by \citet{Houde2009} in the their ADF approximation which also provides an alternative method to derive the dispersion in the field. Therefore and under this consideration, we use the standard error of the mean as:

\begin{equation}
SE = \frac{\sigma}{\sqrt{N}}
\end{equation}

where $\sigma$ is the population standard deviation (unknown but can be estimated from the sample), and $N$ is the population size.
For a 95\% confidence interval (which might be akin to a 95\% credible interval in a Bayesian context, although the interpretation is different), the margin of error is

\begin{equation}
ME = 1.96 \times SE
\end{equation}

If we assume a margin of error to be within 5$^{\circ}$, then $1.96 \times s/\sqrt{N} = 5$ gives the number of independent points required for an error of 5$^{\circ}$, where $s$ is the estimate of $\sigma$. To estimate $s$,  we take the mean of the standard deviation over 
regions in the map where the field appears to be coherent, by visual inspection, on scales of $\sim 1^{\prime \prime}$. This give us an initial population standard deviation estimate of $s \sim 10^{\circ}$. Thus, we obtain $N=16$ independent points as the required number (beams), which when assuming Nyquist sampling for the beam, we obtain a window of $1^{\prime\prime}.5$ in size. The final dispersion map is calculated as:

\begin{equation}
\delta \phi_{i,j} = \sqrt{\sigma_{i,j}^{2} - \sigma_{err, i,j}^{2}} 
\end{equation}

where $\sigma_{err, i,j} = \sigma_{p}/2P_{i,j}$ represents the error in the polarization position angle in an ALMA map, with $\sigma_{P}$ and $P$ being the rms and the polarization intensity respectively \citep{Hull2020}.

\begin{figure*} 
\includegraphics[width=0.95\hsize]{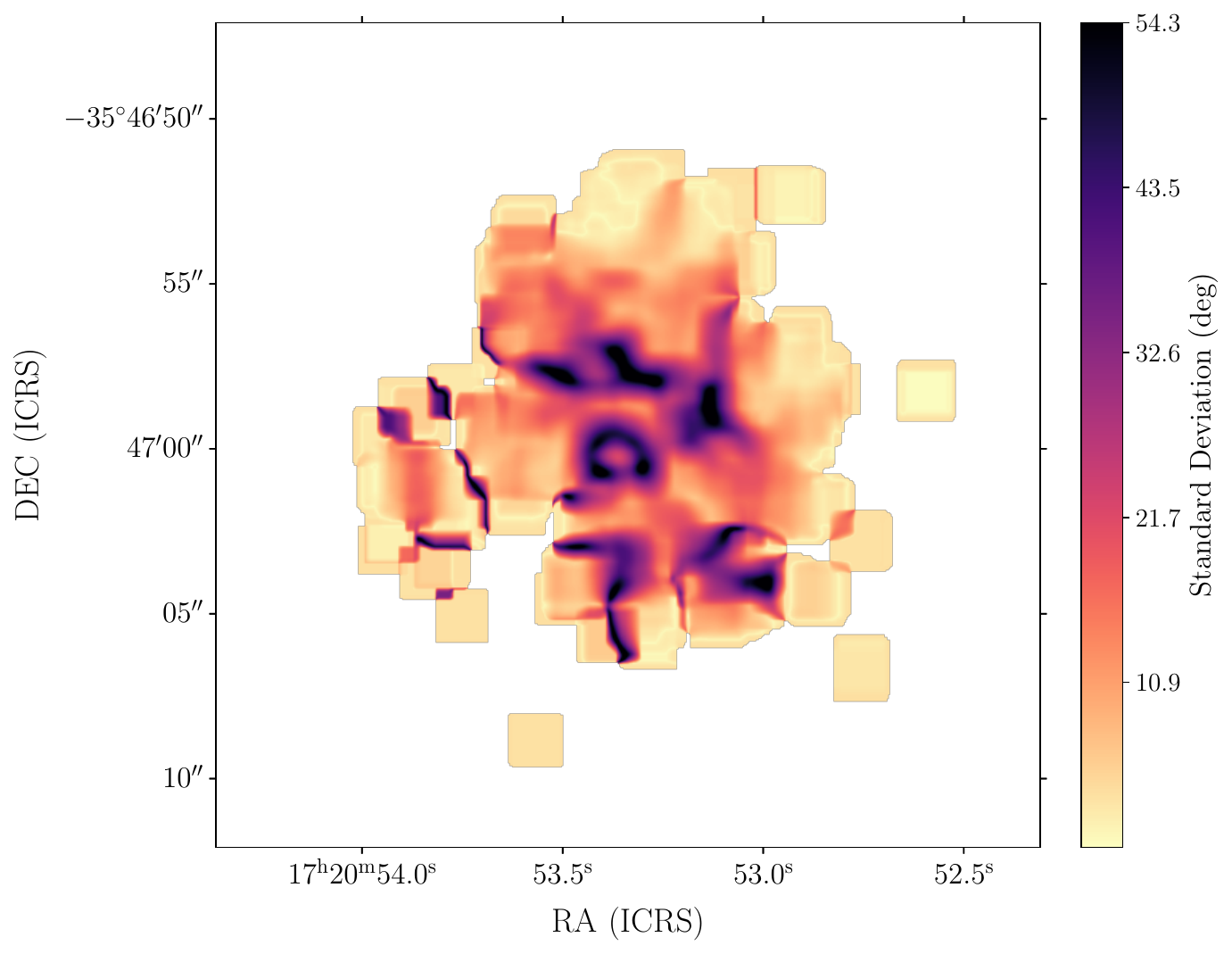}
\smallskip
\caption{ The position angle dispersion map from NGC6334I computed using a moving window of $1^{\prime\prime}.5$ is shown here. The values for the standard deviation are shown in color scale.
\label{fig:std}
}
\bigskip
\end{figure*}

\begin{figure*} 
\includegraphics[width=0.95\hsize]{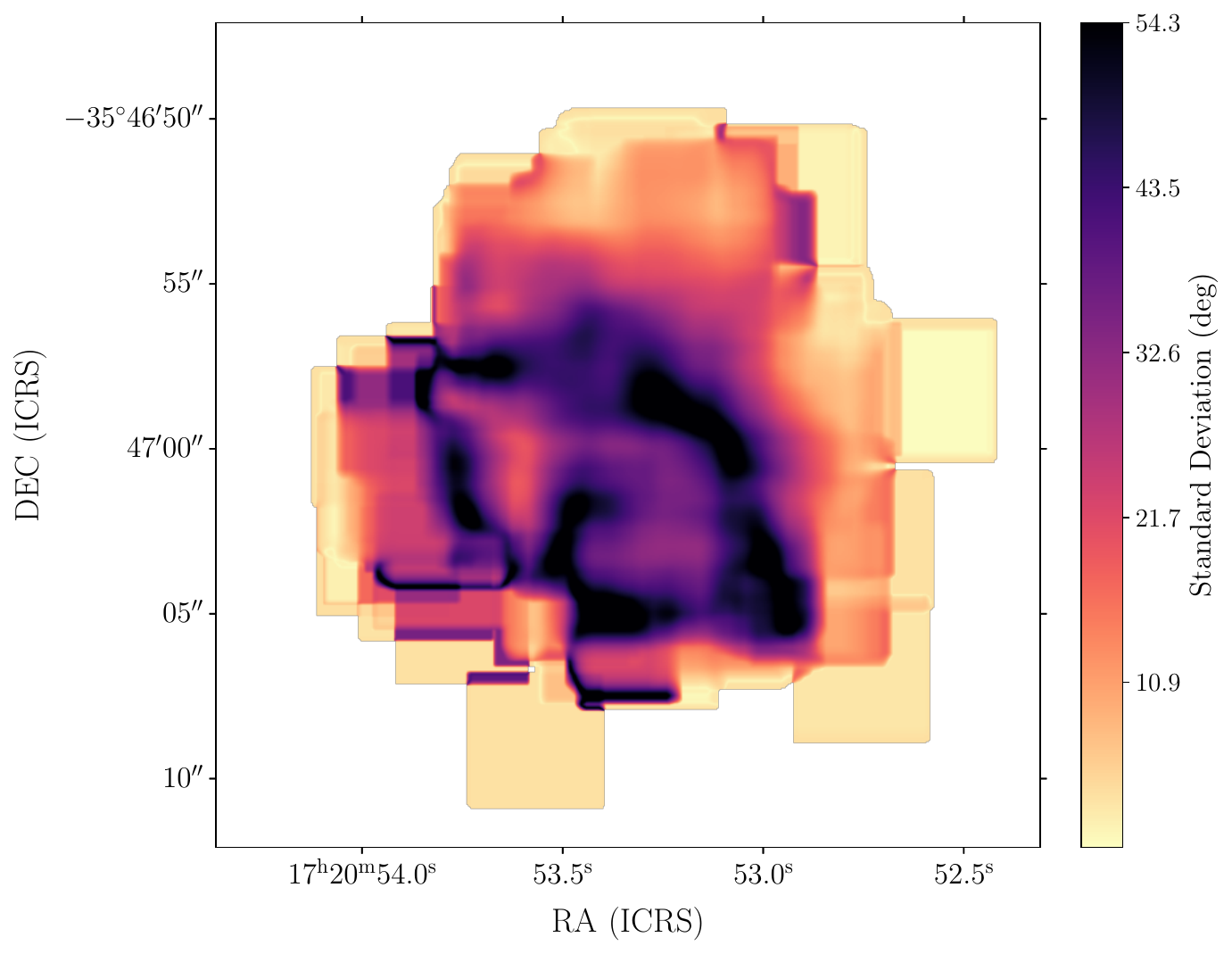}
\smallskip
\caption{ The position angle dispersion map from NGC6334I computed using a moving window of $4^{\prime\prime}$ is shown here. The values for the standard deviation are shown in color scale.
\label{fig:std2}
}
\bigskip
\end{figure*}

Although the choice of $5^{\circ}$ may appear arbitrary, we found it to be a good compromise. A smaller window results in a larger error, while a larger window not only smooths out the map but also biases regions where the field appears smooth. This is because it propagates larger deviations in the field lines seen in the regions where the field has $90^{\circ}$ turns; regions where DCF is likely not fully applicable.
An example of this can be seen in Figure \ref{fig:std2} where we show a dispersion map using a window of 4$^{\prime\prime}$.
 This windows size has an estimated error of $\sim 2^{\circ}$, but the spatial distribution of dispersions values over $25^{\circ}$ is significantly 
larger than our current choice (see Figure \ref{fig:std}). Analysis of the DCF technique done by comparing with simulations, suggest that the applicability of DCF decreases with dispersion values over $25^{\circ}$ \citep[see ][ and references therein ]{Crutcher2004}. Table \ref{tab:std} shows the error estimates for the position angle dispersion for window sizes ranging from $1^{\prime\prime}$ to $7^{\prime\prime}$. 

Finally, it is important to note that  the maximum
value for the standard deviation under semicircular statistics is $\sim 54^{\circ}.45$. This is because, the maximum dispersion is achieved when we have a uniform distribution over the semicircle with a probability density function of $p(r) = 1/\pi$ which yields $\sigma = \sqrt{-2\log{2/\pi}}$.

\setlength{\tabcolsep}{2pt}
\begin{deluxetable*}{ccc}
\tablecolumns{3}
\tablewidth{0pt}
\tabletypesize{\scriptsize}
\tablecaption{ Position Angle Dispersion Error Estimate\label{tab:std}}
\tablehead{
\colhead{Window Size} &
\colhead{Number of Beams} &
\colhead{Dispersion Error} \\
\colhead{$(^{\prime\prime})$}     &
\colhead{} &
\colhead{$(^{\circ})$}     
}
\startdata
1.0 & 6 & 8 \\
1.5 & 16 & 5 \\
2.0 & 25 & 4 \\
2.5 & 40 & 3 \\
3.0 & 56 & 3 \\
4.0 & 100 & 2 \\
5.0 & 156 & 2 \\
6.0 & 225 & 1 \\
7.0 & 306 & 1
\enddata
\tablecomments{The errors where rounded to a single integer for consistency and the number of beams assume Nyquist sampling.
}
\end{deluxetable*}

\section{Simple Error Estimation for DCF}
\label{ap:err_dcf}

Although a number of caveats constrain the interpretation of our result from DCF (see section \ref{sse:caveats}), here we attempt to quatify these uncertainties by employing simple error propagation to estimate an error for
$\mathrm{B}_{\mathrm{pos}}$. 
The expression that we used for DCF is given by equation \ref{eq:Bpos}, which under simple
error propagation becomes,

\begin{equation}
\delta \mathrm{B}_{\mathrm{pos}} = 3\sqrt{\left( \frac{n^{-3/2}\Delta V}{\delta \phi}\sigma_{n} \right)^{2} + \left( \frac{\sqrt{n}}{\delta \phi} \sigma_{\Delta V} \right)^{2} + \left( \frac{\sqrt{n} \Delta V}{\delta \phi^{2}} \sigma_{\delta \phi} \right)^{2}}
\end{equation}

where $\sigma_{n}$, $\sigma_{\Delta V}$, and $\sigma_{\delta \phi}$ are the uncertainties in the number density, velocity dispersion, and position angle dispersion respectively.
The most difficult quantity to estimate is the number density error, because of the uncertainties previously mentioned. We attempt to circumvent this by calculating the mean absolute error between our model and the values derived by \citet{Brogan2016A}. Because both data sets come from ALMA observations from the same source, a number of systematics can be removed by taking the difference between their values and our model. In this way, the mean absolute error can be calculated as,

\begin{equation}
\sigma_{\mathrm{n}} = \frac{\sum^{N}_{i} |n_{i,1} - n_{i,2}|}{N}
\end{equation}

where $n_{i,1}$ corresponds to the number density estimate done by \citet{Brogan2016A} on source $i$, $n_{i,2}$ is the the number density estimate from our model over source $i$, and $N$ is the total number of sources. In this way,  we obtained $\sigma_{\mathrm{n}} = 5\times 10^{7}$ cm$^{-3}$ as the number density error. 
The uncertainty in the velocity dispersion is estimated by the resolution used in the spectral setup of the C$^{33}$S line, or $\sigma_{\Delta V} = 2$ {\kms}, while the error estimate for the position angle dispersion is $\sigma_{\delta \phi} = 5^{\circ}$ (see appendix \ref{ap:std} also for a discussion about the impact a different windows sizes in the error estimate). With all of these values we obtain a mean $\mathrm{B}_{\mathrm{pos}}$ error estimate, $\left< \sigma_{\mathrm{B}} \right> = 1 $ mG after rounding, which yields  $\left< \mathrm{B}_{\mathrm{pos}} \right> = 1.9 \pm 1$ mG. From the values of $\mathrm{B}_{\mathrm{los}}$ obtained by \citet{Hunter2018}, we estimate an error also of 1 mG which after adding them  in quadrature and rounding we obtained  $\left< \mathcal{B} \right>= 4  \pm 1$ mG.

\bigskip

\textit{Facilities:} ALMA.

\textit{Software:}  CASA \citep{CASA2022}.  Astropy \citep{Astropy2018}. MADCUBA \citep{Martin2019}.

\section*{Acknowledgments}
\noindent P.C.C. acknowledges publication  and travel support from ALMA, NAOJ, and NRAO. This work was supported by the NAOJ Research Coordination Committee, NINS (NAOJ-RCC-2202-0401).
P.S. was partially supported by a Grant-in-Aid for Scientific Research (KAKENHI Number JP22H01271 and JP23H01221) of JSPS. 
J.M.G. acknowledges support by the grant PID2020-117710GB-I00 (MCI-AEI-FEDER, UE).  This work is also partially supported by the program Unidad de Excelencia María de Maeztu CEX2020-001058-M.
L.A.Z acknowledges financial support from CONACyT-280775 and UNAM-PAPIITIN110618 grants, Mexico.
A.S.-M.\ acknowledges support from the RyC2021-032892-I grant funded by MCIN/AEI/10.13039/501100011033 and by the European Union `Next GenerationEU’/PRTR, as well as the program Unidad de Excelencia María de Maeztu CEX2020-001058-M, and support from the PID2020-117710GB-I00 (MCI-AEI-FEDER, UE).
This work was supported by the NAOJ Research Coordination Committee, NINS (NAOJ-RCC-2202-0401).
This paper makes use of the following ALMA data: ADS/JAO.ALMA\#2018.1.00105.S.
ALMA is a partnership of ESO (representing its member states), NSF (USA) and NINS (Japan), together with NRC (Canada), MOST and ASIAA (Taiwan), and KASI (Republic of Korea), in cooperation with the Republic of Chile. The Joint ALMA Observatory is operated by ESO, AUI/NRAO and NAOJ.
The National Radio Astronomy Observatory is a facility of the National Science Foundation operated under cooperative agreement by Associated Universities, Inc.

\bibliography{biblio}
\bibliographystyle{aasjournal}

\end{document}